\pdfoutput=1
\documentclass{article}

\usepackage{arxiv}
\usepackage{amsmath,amssymb,amsfonts,amsthm,mathtools}
\usepackage[utf8]{inputenc}
\usepackage[T1]{fontenc}
\usepackage[round]{natbib}
\usepackage{url}
\usepackage{booktabs}
\usepackage{array}
\usepackage[flushleft]{threeparttable}
\usepackage{microtype}
\usepackage{xcolor}
\usepackage{graphicx}
\usepackage{caption}
\usepackage{algorithm}
\usepackage{algorithmic}
\usepackage{enumitem}
\usepackage[section]{placeins}
\usepackage{hyperref}
\hypersetup{colorlinks=true,
  linkcolor={black!45!blue}, citecolor={black!45!blue}, urlcolor={black!45!blue},
  pdftitle={Meaning-Dependent Judge Sensitivity Changes the Minimax Rate of Semantic Entropy},
  pdfauthor={Xiaoli Yu; Jiamiao Liu},
  pdfkeywords={entropy estimation, large language models, method of moments, minimax rates, uncertainty quantification}}
\usepackage[capitalize,noabbrev]{cleveref}

\theoremstyle{plain}
\newtheorem{theorem}{Theorem}
\newtheorem{corollary}{Corollary}
\newtheorem{proposition}{Proposition}
\newtheorem{lemma}{Lemma}
\newtheorem{condition}{Condition}

\crefname{condition}{Condition}{Conditions}
\Crefname{condition}{Condition}{Conditions}

\newcommand{\Real}{\mathbb{R}}
\newcommand{\Cpx}{\mathbb{C}}
\newcommand{\Exp}{\mathbb{E}}

\newcommand{\Var}{\mathrm{Var}}
\newcommand{\Cov}{\mathrm{Cov}}
\newcommand{\Bern}{\mathrm{Bern}}
\newcommand{\TV}{\mathrm{TV}}
\newcommand{\Ind}[1]{\mathbf{1}\{#1\}}
\newcommand{\Ent}{H}
\newcommand{\hH}{\widehat{H}}
\newcommand{\Simplex}{\Theta_K}
\newcommand{\htheta}{\widehat{\theta}}
\newcommand{\hS}{\widehat{S}}
\newcommand{\eps}{\varepsilon}
\newcommand{\SG}{S_{\mathrm{G}}}
\newcommand{\Sobs}{S_{\mathrm{obs}}}
\newcommand{\Sw}{\mathrm{Sw}}
\newcommand{\Pu}{\mathsf{P}}
\newcommand{\DS}{\mathsf{D}}
\newcommand{\Rem}{\mathrm{Rem}}

\newcommand{\gap}{\gamma}
\newcommand{\one}{\mathbf{1}}

\graphicspath{{figures/}}

\renewcommand{\shorttitle}{Judge Sensitivity and Semantic Entropy Rates}

\title{Meaning-Dependent Judge Sensitivity Changes the Minimax Rate of Semantic Entropy}

\author{%
  Xiaoli Yu \\
  School of Cyber Security and Information Law \\
  Chongqing University of Posts and Telecommunications \\
  Chongqing 400065, China
  \And
  Jiamiao Liu\thanks{Corresponding author.} \\
  Department of Information, Xinqiao Hospital \\
  Army Medical University (Third Military Medical University) \\
  Chongqing 400037, China \\
  \texttt{ljm6046@163.com}
}

\date{}

\begin{document}

\maketitle

\begin{abstract}
Semantic entropy, the entropy of the meanings among a language model's sampled answers, is estimated from
noisy pairwise judgments. Its minimax theory assumes a judge that recognizes every meaning equally well. We show
that a meaning-dependent sensitivity changes the minimax rate even when it is known exactly and the judge
recognizes every meaning at least as well as a homogeneous one. Two answers with meaning $c$ are declared
equivalent with probability $q+\Delta w_c$, two with different meanings with probability $q$, where
$q$ is bounded away from $0$ and $1$ and the number $K\ge3$ of meanings is fixed. For almost every sensitivity
profile $w$, the entropy is uniformly consistently estimable exactly when $N:=n^{2K-1}\Delta^{4K-4}\to\infty$: for
$\Delta=n^{-\beta}$ the threshold falls from $\beta<K/(2K-2)$ for a homogeneous judge to
$\beta<(2K-1)/(4K-4)$, that of such a judge on $2K-1$ meanings, and from $3/4$ to $5/8$ for three
meanings. On every fixed polynomial window $n\Delta^2\le n^{-\kappa}$ the minimax squared risk is of order
$\min\{1,N^{-1/(2K-1)}\}$. Star statistics computable in $O(n^2K)$ time and a projection attain this rate
through a weighted entropy modulus with optimal exponent $1/(2K-1)$. The lower bound uses a moment-preserving
path near the interior point $\theta\propto1/w$, along which the judge's channel moves with the distribution.

\end{abstract}


\section{Introduction}
\label{sec:intro}

Semantic entropy, the entropy of the meanings among a language model's sampled answers, measures the model's
uncertainty on a prompt and is estimated from noisy pairwise judgments of whether two answers share a
meaning \citep{kuhn2023semantic,farquhar2024detecting}. The minimax theory of this estimation problem,
developed in concurrent work \citep{anonymous2026four,anonymous2026three,anonymous2026rarity}, assumes a judge
that recognizes every meaning equally well. The reliability of automatic answer judges, however, varies with the
form of the answers, for instance with their length \citep{wang2023evaluating,kamalloo2023evaluating}. If the
answers expressing one meaning are short and those expressing another are long, the probability that the
judge recognizes two paraphrases as equivalent depends on the meaning. We show that such a dependence changes
the minimax rate even when it is known exactly; for three meanings, almost every sensitivity profile and $\Delta=n^{-\beta}$, the consistency
threshold falls from $\beta<3/4$ to $\beta<5/8$, even for a judge that recognizes every meaning at least as well as a
homogeneous one.

\begin{figure}[t]
  \centering
  \includegraphics{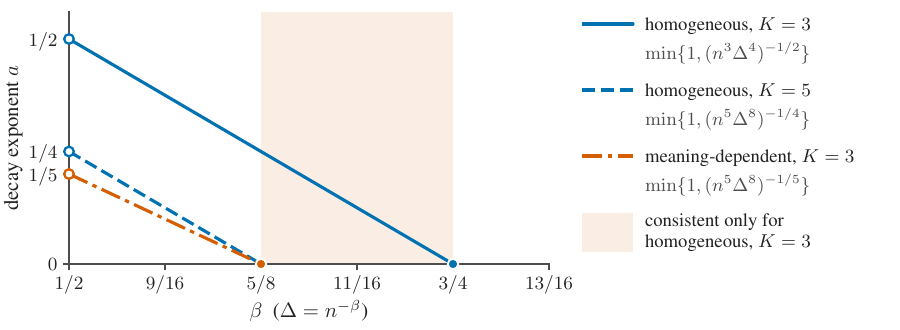}
  \caption{Polynomial decay exponent $a$ of the minimax squared risk of semantic entropy, $R^*=n^{-a+o(1)}$,
  when $\Delta=n^{-\beta}$ and $q\in[q_0,1-q_0]$ is fixed. Blue: a homogeneous judge, equally sensitive to
  every meaning, on $K=3$ (solid) and $K=5$ (dashed) meanings (\citealp[Cor.~3]{anonymous2026three};
  \citealp[Cor.~1]{anonymous2026rarity}). Orange (dash-dotted): a meaning-dependent judge on $K=3$ meanings with a
  fixed sensitivity profile outside a Lebesgue-null set (\Cref{thm:main}). Beyond its threshold each risk
  is of constant order. The rates are matched on every fixed polynomial window, and these windows cover every fixed
  $\beta>1/2$; open circles mark the excluded endpoint $\beta=1/2$, where $n\Delta^2=1$. In the shaded band
  the homogeneous judge on three meanings allows uniformly consistent estimation and the meaning-dependent
  judge does not; the meaning-dependent judge on three meanings has the threshold of the homogeneous judge
  on five.}
  \label{fig:hero}
\end{figure}

We model a judge whose sensitivity is known and depends on the meaning: two answers with meaning $c$ are
declared equivalent with probability $q+\Delta w_c$, two answers with different meanings with probability $q$,
and the sensitivity profile $w=(w_1,\dots,w_K)$ is part of the known channel; $w\equiv1$ is the homogeneous judge
of the existing theory. We study the weak regime $n\Delta^2\to0$, where, for a homogeneous judge on three meanings, no
estimator of the meanings beats the majority rule \citep{anonymous2026three}, so the entropy has to be estimated
without recovering the meanings. For $K\ge3$, $q$ bounded away from $0$ and $1$, and almost every $w$, the minimax squared
risk of estimating the entropy from the verdicts on every fixed polynomial window $n\Delta^2\le n^{-\kappa}$ is of order
\[
  \min\big\{1,N_{2K-1}^{-1/(2K-1)}\big\},\qquad N_{2K-1}:=n^{2K-1}\Delta^{4K-4}
\]
(\Cref{thm:main,prop:generic}), and the entropy is uniformly consistently estimable exactly when
$N_{2K-1}\to\infty$ (\Cref{cor:threshold}): for $\Delta=n^{-\beta}$ the threshold drops from the homogeneous
$K/(2K-2)$ \citep[Cor.~1]{anonymous2026rarity} to $(2K-1)/(4K-4)$, that of a homogeneous judge on $2K-1$ meanings
(\Cref{fig:hero}, \Cref{tab:comparison}). The slow rate is attained by distributions whose meanings all have
probability bounded away from zero, whereas for a homogeneous judge the risk is set by distributions with rare
meanings near a vertex of the simplex.

\begin{table}[t]
  \centering
  \begin{threeparttable}
    \caption{Minimax squared risk of semantic entropy from a judge with known channel parameters.}
    \label{tab:comparison}
    \footnotesize
    \begin{tabular}{@{}>{\raggedright\arraybackslash}p{0.32\textwidth}lc>{\raggedright\arraybackslash}p{0.14\textwidth}>{\centering\arraybackslash}p{0.10\textwidth}@{}}
\toprule
setting & minimax risk & $\beta_c$ & least-favorable family & matched moments \\
\midrule
meanings observed (no judge) & $1/n$ & none & regular interior & none \\
homogeneous judge, at most three meanings, window $n\Delta^2\le c_1$ \citep[Cor.~3]{anonymous2026three} & $\min\{1,N_3^{-1/2}\}$ & $3/4$ & near a vertex & $2$ \\
homogeneous judge, $K\ge3$ meanings, windows $n\Delta^2\le n^{-\kappa}$ \citep[Cor.~1]{anonymous2026rarity} & $\min\{1,N_K^{-1/(K-1)}\}$ & $\frac{K}{2K-2}$ & near a vertex & $K-1$ \\
meaning-dependent judge, $K\ge3$ meanings, generic $w$, windows $n\Delta^2\le n^{-\kappa}$ (\Cref{thm:main}) & $\min\{1,N_{2K-1}^{-1/(2K-1)}\}$ & $\frac{2K-1}{4K-4}$ & near $\theta\propto1/w$ & $2K-2$ \\
\bottomrule
\end{tabular}

    \begin{tablenotes}[flushleft]\footnotesize
    \item[]
      The risk is that of $n$ sampled answers with $q\in[q_0,1-q_0]$, up to constants depending on the fixed
      parameters; $N_v:=n^v\Delta^{2v-2}$. The matching lower bounds hold on the stated windows, with $c_1>0$ and
      fixed $\kappa\in(0,1)$; the threshold $\beta_c$ refers to $\Delta=n^{-\beta}$. The last two columns give the
      location of a least-favorable family and the number of leading weighted tree moments it matches
      ($P_1,\dots,P_{K-1}$ for a homogeneous judge, $S_1,\dots,S_{2K-2}$ for a meaning-dependent one).
    \end{tablenotes}
  \end{threeparttable}
\end{table}

\subsection{Why the Rate Changes}
\begin{figure}[t]
  \centering
  \includegraphics{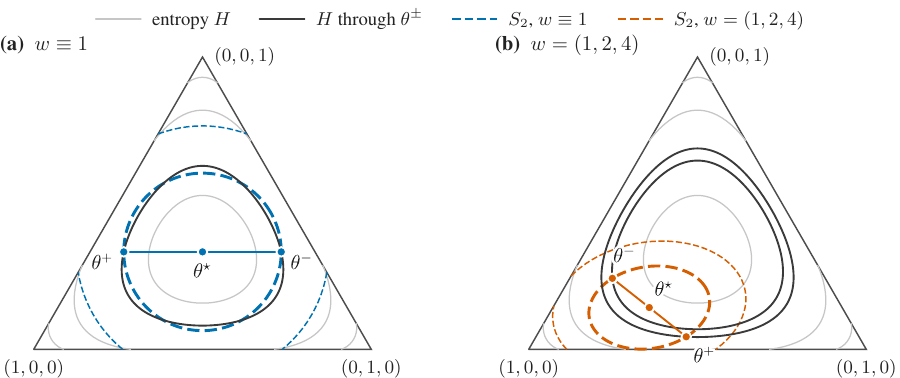}
  \caption{Entropy contours (gray) and contours of the second weighted tree moment $S_2$ (dashed) on the simplex
  of distributions on three meanings; the dark curves are the entropy contours through $\theta^\pm$ and the
  thick dashed curve is the $S_2$ contour through both.
  (a)~Homogeneous judge, $w\equiv1$: the $S_2$ circles are centered at $\theta^\star=(1/3,1/3,1/3)$, where the
  entropy is maximal, and the pair $\theta^\pm=(1/3\pm s,1/3\mp s,1/3)$ lies on one entropy contour.
  (b)~$w=(1,2,4)$: the $S_2$ ellipses are centered at $\theta^\star=(4,2,1)/7$, where all locations $w_c\theta_c$
  equal $\mu=4/7$. The pair $\theta^\pm_c=a_c(\mu\pm sx_c)$, with $x\approx(-0.149,0.797,-1)$ solving
  $\sum_ca_cx_c=\sum_ca_cx_c^3=0$, shares $S_1,\dots,S_4$ and lies on the entropy contours $H=0.838$ and
  $H=0.896$. Both pairs have $s=0.7\mu$, so $s=0.4$ in (b).}
  \label{fig:simplex}
\end{figure}

Knowing $w$ does not tell which meaning an answer has, so no single verdict can be corrected for its sensitivity.
The judge reveals the meanings only through moments. With $X_{ij}:=(A_{ij}-q)/\Delta$, the product of the $X_e$
along any tree on $v$ answers has mean $S_v(\theta)=\sum_cw_c^{v-1}\theta_c^v$, the $v$-th moment of the measure
$\nu_\theta=\sum_ca_c\delta_{w_c\theta_c}$ with known masses $a_c=1/w_c$ and unknown locations $w_c\theta_c$. At the
interior point $\theta^\star\propto1/w$ all $K$ locations equal $\mu:=1/\sum_ca_c$. Moving them to $\mu\pm sx_c$, where $x$ solves
$\sum_ca_cx_c^j=0$ for every odd $j\le2K-3$, gives two distributions whose first $2K-2$ weighted tree moments
agree and whose $(2K-1)$-th differ by order $s^{2K-1}$, while their entropies differ by
$2s\ell(x)+O(s^3)$ with $\ell(x)=-\sum_ca_cx_c\log a_c$; for almost every $w$ some such $x$ has $\ell(x)\ne0$ (\Cref{fig:simplex}, panel~b). For a homogeneous
judge, $\theta^\star$ is the uniform distribution, where the entropy is stationary (\Cref{fig:simplex}, panel~a), and the least-favorable
distributions sit near the vertices and match only $K-1$ moments. Weighted tree moments determine the rate;
cyclic motifs carry further information, and the lower bound controls it.

\subsection{Contributions}
\begin{enumerate}[leftmargin=*,itemsep=1pt,topsep=2pt]
  \item The minimax rate on every fixed polynomial window and the consistency threshold of semantic entropy under a judge with
    known meaning-dependent sensitivity, for every $K\ge3$ and almost every sensitivity profile, with an explicit
    criterion for $K=3$ and a least-favorable family at the interior point $\theta^\star\propto1/w$ instead of near a
    vertex of the simplex (\Cref{thm:main}, \Cref{cor:threshold}, \Cref{prop:generic}).
  \item A global weighted entropy modulus with the optimal exponent $1/(2K-1)$ (\Cref{thm:modulus}). A
    projected moment estimate may land anywhere in the simplex, the entropy has a logarithmic singularity at
    its boundary, and the moments of $\nu_\theta$ do not say which meaning sits at which location. The modulus
    handles all three, the last through the subset-sum gap of the masses $a_c$. With it, star statistics
    computed in $O(n^2K)$ operations and a projection solved on a grid of polynomial size attain the rate
    (\Cref{cor:projection}).
  \item A lower bound for experiments whose channel must move. With $w$ fixed, the first $K$ weighted tree
    moments determine $\theta$ near the least-favorable distributions, so no path between them keeps the
    required $2K-2$ moments fixed. Instead, the sensitivities move along a path of Gauss-quadrature measures
    (\Cref{fig:path}), and a sweep inequality bounds the Fisher information of the verdict graph with an
    expansion in which the channel varies (\Cref{sec:lower}).
\end{enumerate}

\subsection{Organization}
\Cref{sec:setup} sets up the model, \Cref{sec:results} states the results, \Cref{sec:upper,sec:lower}
describe the upper and the lower bound, \Cref{sec:related} discusses related work, and \Cref{sec:conclusion}
concludes. Every result is proved in the appendices with its hypotheses restated: the tools in
\Cref{app:tools}, the weighted entropy modulus in \Cref{app:modulus}, the moment statistics and the upper bound
in \Cref{app:upper}, the path in \Cref{app:path}, the information inequality in \Cref{app:information}, the
lower bound and \Cref{cor:threshold} in \Cref{app:lower}, and genericity in \Cref{app:generic}.
\Cref{app:numerics} describes the numerical checks, with their settings and arithmetic precision.
The code behind every computed number in the paper, the outputs that hold these numbers, and instructions that
reproduce them are available in the public repository \url{https://github.com/yahiko-l/semantic-entropy-minimax}.

\section{Setup}
\label{sec:setup}

We define the model, the weighted tree moments and the witness condition.

\subsection{Model}
A language model is sampled $n$ times on one prompt. Answer $i$ has a meaning $Z_i\in[K]:=\{1,\dots,K\}$, and
$Z_1,\dots,Z_n$ are i.i.d.\ from $\theta\in\Simplex:=\{\theta\in[0,1]^K:\sum_c\theta_c=1\}$, where $K\ge3$ is known and
zero coordinates are allowed. The target is the entropy $\Ent(\theta):=-\sum_c\theta_c\log\theta_c$. A judge compares
every pair of answers and reports $A_{ij}\in\{0,1\}$; given the meanings, the verdicts are independent, with
\begin{equation}
  \label{eq:model}
  A_{ij}\mid Z\ \sim\ \Bern\big(q+\Delta\,w_{Z_i}\Ind{Z_i=Z_j}\big),\qquad1\le i<j\le n .
\end{equation}
The channel $(q,\Delta,w)$ is known, with $q\ge0$, $\Delta>0$, sensitivities $w\in(0,\infty)^K$ and
$q+\Delta U\le1$, $U:=\max_cw_c$. Rescaling $(\Delta,w)\mapsto(\Delta/\lambda,\lambda w)$ leaves the model
unchanged, and our statements are for fixed $w$; $w\equiv1$ is the homogeneous judge. We write $m:=2K-1$,
$\delta:=n\Delta^2$ and $N_v:=n\delta^{v-1}=n^v\Delta^{2v-2}$. The weak regime is $\delta\to0$, and a polynomial window
is $\{\delta\le n^{-\kappa}\}$ for a fixed $\kappa\in(0,1)$; for $\Delta=n^{-\beta}$ these windows cover every fixed
$\beta>1/2$. The minimax risk is $R^*_n(w):=\inf_{\hH}\sup_{\theta\in\Simplex}\Exp_\theta(\hH-\Ent(\theta))^2$, where the infimum is over all
estimators $\hH$ computed from the verdicts $(A_{ij})_{i<j}$, the known channel and auxiliary randomness. For positive $a,b$ we write $a\lesssim b$ if $a\le Cb$ and
$a\asymp b$ if $a\lesssim b\lesssim a$, where $C$ does not depend on $n$, $q$, $\Delta$, $\theta$ or the amplitude $s$
used below and may depend on $K$, $w$, $q_0$, $\kappa$ and the witness $x$; $O(\cdot)$ is used in the same sense.

\subsection{Weighted Tree Moments}
Let $X_{ij}:=(A_{ij}-q)/\Delta$, so that $\Exp[X_{ij}\mid Z]=w_{Z_i}\Ind{Z_i=Z_j}$. For any tree on $v$ distinct
answers, the product of the $X_e$ along its edges has mean
\[
  S_v(\theta):=\sum_cw_c^{v-1}\theta_c^v=\int x^v\,\mathrm d\nu_\theta(x),\qquad\nu_\theta:=\sum_ca_c\delta_{u_c},\quad
  a_c:=1/w_c,\quad u_c:=w_c\theta_c ,
\]
a sensitivity-weighted version of the probability $\sum_c\theta_c^v$ that the $v$ answers share a meaning. The
weighted tree moments are thus the moments of a measure with known masses $a_c$ and unknown locations
$u_c$, and $S_1\equiv1$. For
$w\equiv1$ they are the power sums $P_v(\theta):=\sum_c\theta_c^v$ of the homogeneous analyses. We write $\|a\|_1:=\sum_ca_c$,
$\mu:=1/\|a\|_1$ and $\theta^\star:=\mu a$, the point at which all locations equal $\mu$. The product $\odot$ is entrywise, and $\one$ is the all-ones vector.

\subsection{Gap and Witness}
The subset-sum gap of the masses is
\[
  \gap(w):=\min\Big\{\Big|\sum_{c\in I}a_c-\sum_{c\in J}a_c\Big|:\ I,J\subseteq[K]\ \text{disjoint, not both empty}\Big\}.
\]
It is positive unless two disjoint groups of reciprocal sensitivities have equal sums; in particular it
vanishes when two sensitivities coincide, so the homogeneous judge, which has a different rate, is excluded.
For $x\in\Real^K$ let $X_j(x):=\sum_ca_cx_c^j$ and $\ell(x):=-\sum_ca_cx_c\log a_c$.

\begin{condition}[Witness]
  \label{cond:W}
  There is $x\in\Real^K$ with pairwise distinct coordinates such that $X_j(x)=0$ for every odd $j\le2K-3$
  and $\ell(x)\ne0$.
\end{condition}

A witness gives the two distributions $\theta^\pm:=a\odot(\mu\one\pm sx)$, whose weighted tree moments agree up
to order $2K-2$; $\ell(x)$ is the derivative of the entropy along $s\mapsto\theta^+$ at $s=0$.

\section{Main Results}
\label{sec:results}

\begin{theorem}[Minimax Rate]
  \label{thm:main}
  Fix $K\ge3$ and $w\in(0,\infty)^K$, and let $m=2K-1$.
  \begin{enumerate}[label=(\roman*),leftmargin=*]
    \item If $\gap(w)>0$, there is $C_U$ depending only on $(K,w)$ such that for every $n\ge2m$ and every
      $q\ge0$, $\Delta>0$ with $q+\Delta U\le1$, the estimator $\hH$ of \Cref{alg:main} satisfies
      \[
        \sup_{\theta\in\Simplex}\Exp_\theta(\hH-\Ent(\theta))^2\le\min\big\{(\log K)^2,\ C_U\max\{1/n,1/N_m\}^{1/m}\big\}.
      \]
    \item If $w$ satisfies \Cref{cond:W}, then for every $q_0\in(0,1/2]$ and $\kappa\in(0,1)$ there are $c_L>0$
      and $n_L$ depending only on $(K,w,q_0,\kappa)$ such that $R^*_n(w)\ge c_L\min\{1,N_m^{-1/m}\}$ whenever
      $n\ge n_L$, $q\in[q_0,1-q_0]$ and $n\Delta^2\le n^{-\kappa}$.
  \end{enumerate}
  In particular, if $\gap(w)>0$ and \Cref{cond:W} holds, then $R^*_n(w)\asymp\min\{1,N_m^{-1/m}\}$ on every fixed
  polynomial window, uniformly over $q\in[q_0,1-q_0]$.
\end{theorem}

Part (i) holds for every channel and every $n\Delta^2$; for $n\Delta^2\le1$ we have $N_m\le n$, and the bound reads
$C_UN_m^{-1/m}$. Part (ii) uses only \Cref{cond:W}, so it also covers a profile with tied sensitivities, for which
$\gap(w)=0$, whenever \Cref{cond:W} holds, as it does for $w=(1,1,2)$ with $x=((1+\sqrt{5})/2,(1-\sqrt{5})/2,-2)$.
Both parts are proved in \Cref{sec:upper,sec:lower}, with
full proofs in \Cref{app:upper,app:lower}.

\begin{corollary}[Consistency Threshold]
  \label{cor:threshold}
  Let $\gap(w)>0$ and let \Cref{cond:W} hold. For any sequence $(q_n,\Delta_n)$ with $q_n\in[q_0,1-q_0]$ and
  $q_n+\Delta_nU\le1$, estimators with $\sup_\theta\Exp_\theta(\hH_n-\Ent(\theta))^2\to0$ exist if and only if
  $n^{2K-1}\Delta_n^{4K-4}\to\infty$; no window condition is needed. For $\Delta_n=n^{-\beta}$ this holds exactly when
  $\beta<(2K-1)/(4K-4)$, which is the threshold $K'/(2K'-2)$ of a homogeneous judge on $K'=2K-1$ meanings and
  lies $1/(4K-4)$ below the homogeneous threshold $K/(2K-2)$ on $K$ meanings.
\end{corollary}

No window condition is needed because \Cref{thm:main}(i) holds for every $n\Delta^2$, while a bounded $N_m$ forces
$n\Delta^2\le n^{-1/(4K-4)}$ for large $n$, which is the polynomial window with $\kappa=1/(4K-4)$ (\Cref{app:lower}).

\subsection{The Homogeneous Baseline}
For a judge that treats all meanings alike, $w\equiv1$, and $q\in[q_0,1-q_0]$, the homogeneous analyses give the following. For at most three meanings the minimax squared risk is of order $\min\{1,N_3^{-1/2}\}$ on the whole
weak window $n\Delta^2\le c_1$ \citep[Cor.~3]{anonymous2026three}; for four meanings it is of order
$\min\{1,N_4^{-1/3}\}$ on every fixed polynomial window \citep[Thms.~1--2]{anonymous2026four}; for fixed $K\ge3$ it
is of order $\min\{1,N_K^{-1/(K-1)}\}$ on every fixed polynomial window, and the entropy is uniformly consistently
estimable exactly when $N_K\to\infty$, with no window condition \citep[Thm.~1, Cor.~1]{anonymous2026rarity}. In
the general analysis the least-favorable distributions lie near a vertex of the simplex and match
$P_1,\dots,P_{K-1}$. A meaning-dependent judge on $K$ meanings thus has the threshold of a homogeneous judge on
$2K-1$ meanings. Both risks are powers of the same $N_{2K-1}$, with exponent $1/(2K-1)$ for the meaning-dependent
judge on $K$ meanings and $1/(2K-2)$ for the homogeneous judge on $2K-1$ meanings.

\subsection{Genericity of the Hypotheses}
The hypotheses of \Cref{thm:main} hold for almost every sensitivity profile.

\begin{proposition}[Genericity]
  \label{prop:generic}
  Let $K\ge3$. If $\gap(w)>0$, every nonzero real solution of $X_1(x)=X_3(x)=\dots=X_{2K-3}(x)=0$ has pairwise
  distinct coordinates, so \Cref{cond:W} holds if and only if $\ell(x)\ne0$ for some such solution; nonzero real
  solutions always exist by the Borsuk--Ulam theorem. The set of $w\in(0,\infty)^K$ for which $\gap(w)=0$ or
  \Cref{cond:W} fails is Lebesgue-null. For $K=3$, $\gap(w)>0$ and
  \[
    g(w):=w_1^2\log^3\frac{w_2}{w_3}+w_2^2\log^3\frac{w_3}{w_1}+w_3^2\log^3\frac{w_1}{w_2}\ne0
  \]
  imply \Cref{cond:W}; for example $w=(1,2,4)$ has $\gap(w)=1/4$ and $g(w)=15\log^32$.
\end{proposition}

The proof (\Cref{app:generic}) writes $a=t\alpha(x)$, with $t>0$ and an explicit rational map $\alpha$, whenever $\gap(w)>0$ and $x$ is a nonzero real solution rescaled to unit length with a suitable sign. Then $\ell(x)$ becomes $-t$ times a real-analytic function of $x$ that is not identically zero. \Cref{thm:main,prop:generic} together give the rate
for every $K\ge3$ and almost every $w$. A judge whose profile has every $w_c\ge1$ recognizes every meaning at least as
well as the homogeneous judge with the same $q$ and $\Delta$. Such profiles form a set of positive Lebesgue measure, so
by \Cref{prop:generic} almost all of them satisfy the hypotheses of \Cref{thm:main}, and by \Cref{cor:threshold} their
threshold is $\beta<(2K-1)/(4K-4)$ instead of $\beta<K/(2K-2)$, for example $\beta<5/8$ instead of $\beta<3/4$ for
$w=(1,2,4)$.

\subsection{Where the Difficulty Sits}
A least-favorable family lies near $\theta^\star=\mu a$, at which every meaning has probability bounded
away from zero: the pair $\theta^\pm=a\odot(\mu\one\pm sx)$ shares $S_1,\dots,S_{2K-2}$, its $S_{2K-1}$ differ by
$2s^{2K-1}|X_{2K-1}(x)|$, and its entropies differ by $2s\ell(x)+O(s^3)$. The amplitude
$s=\min\{s_1,a_0N_m^{-1/(2m)}\}$, with constants $s_1,a_0>0$, makes the two laws of the verdicts indistinguishable and gives the
rate, including its saturation at one when $N_m$ is bounded.

\section{The Upper Bound}
\label{sec:upper}

The estimator averages products of centered verdicts along stars, projects the averages onto the set of
weighted tree moments, and reads off the entropy.

\begin{algorithm}[!t]
  \caption{Projected Star-Moment Estimator}
  \label{alg:main}
  \begin{algorithmic}[1]
    \REQUIRE verdicts $(A_{ij})_{i<j}$, channel $(q,\Delta,w)$, number $K$ of meanings; $m=2K-1$
    \STATE $X_{ij},X_{ji}\leftarrow(A_{ij}-q)/\Delta$ for $i<j$;\quad $T_2,\dots,T_m\leftarrow0$
    \FOR{$i=1,\dots,n$}
      \STATE $(e_0,e_1,\dots,e_{m-1})\leftarrow(1,0,\dots,0)$
      \FORALL{$j\ne i$}
        \STATE $e_k\leftarrow e_k+X_{ij}e_{k-1}$ for $k=m-1,m-2,\dots,1$, in this order
      \ENDFOR
      \STATE $T_v\leftarrow T_v+e_{v-1}$ for $v=2,\dots,m$
    \ENDFOR
    \STATE $\hS_v\leftarrow(v-1)!\,T_v/(n)_v$ for $v=2,\dots,m$
    \STATE $\htheta\leftarrow$ a minimizer of $\max_{2\le v\le m}|S_v(\vartheta)-\hS_v|$ over $\vartheta\in\Simplex$
    \RETURN $\hH:=\Ent(\htheta)$
  \end{algorithmic}
\end{algorithm}

After the loop over $j$, $e_k$ is the elementary symmetric polynomial of degree $k$ in $(X_{ij})_{j\ne i}$, and
$(n)_v=n(n-1)\cdots(n-v+1)$; the statistic $\hS_v$ is the average of $\prod_kX_{ij_k}$ over all stars with center $i$
and distinct leaves $j_1,\dots,j_{v-1}$. \Cref{cor:projection} gives the same rate for a minimizer over a grid of
polynomial size.

\begin{lemma}[Star Statistics]
  \label{lem:var}
  For $2\le v\le m$, $\Exp\hS_v=S_v(\theta)$, and for $n\ge2v-2$
  \[
    \Var(\hS_v)\le B_v\max\{1/n,1/\tilde N_v\},\qquad\tilde N_v:=\frac{n^v}{(U^2+\rho)^{v-1}},
  \]
  where $\rho:=\max\{q(1-q),\max_cp_c(1-p_c)\}/\Delta^2$, $p_c=q+\Delta w_c$ and $B_v:=2^{2v+1}v!\max\{1,U\}^{2v-2}$. The same holds for the average over the
  embeddings of any tree on $v$ vertices. \Cref{alg:main} computes $\hS_2,\dots,\hS_m$ in $O(n^2m)$ operations.
\end{lemma}

Embeddings with disjoint vertex sets are independent, and two embeddings that share $s$ vertices share
at most $s-1$ edges, because the shared edges form a forest; counting the pairs gives the bound
(\Cref{app:upper}). Since $q+\Delta U\le1$ implies $\Delta^2(U^2+\rho)\le5/4$, we have $\tilde N_v\ge(4/5)^{v-1}N_v$.

On the weak window, $N_v$ decreases in $v$, so the variance bound of \Cref{lem:var} is largest for the top statistic $\hS_m$. The
difficulty is to convert moment accuracy into entropy accuracy uniformly over the simplex.

\begin{theorem}[Global Weighted Entropy Modulus]
  \label{thm:modulus}
  Let $\gap(w)>0$ and, for $\theta,\theta'\in\Simplex$, $\epsilon:=\max_{2\le v\le m}|S_v(\theta)-S_v(\theta')|$.
  \begin{enumerate}[label=(\alph*),leftmargin=*]
    \item If $\epsilon\le1$, then $\max_c|w_c\theta_c-w_c\theta'_c|\le(C_1/\gap(w))\,\epsilon^{1/m}$ with
      $C_1:=(1+U)^{2K-2}/2+{(4K-4)\|a\|_1}$.
    \item $|\Ent(\theta)-\Ent(\theta')|\le C_K(w)\,\epsilon^{1/m}$, with $C_K(w)$ given in \eqref{eq:b:CK} and depending
      only on $K$, $U$, $\min_cw_c$, $\|a\|_1$ and $\gap(w)$.
  \end{enumerate}
\end{theorem}

\emph{Proof idea} (\Cref{app:modulus}). The measures $\nu_\theta$ and $\nu_{\theta'}$ have total mass $\|a\|_1$,
support in $[0,U]$ and at most $2K$ atoms together. A polynomial with at most $2K-2$ roots that has the sign
of the difference of their distribution functions turns the moment gap into a bound $W_1(\nu_\theta,\nu_{\theta'})
\lesssim\epsilon^{1/m}$. This step follows the moment comparison of \citet{wu2020optimal}, which uses a sign-preserving
polynomial in the spirit of \citet{moitra2010settling}. Two steps are new. First, wherever the classes below a point
differ for the two measures, their distribution functions differ by a signed subset sum of the masses, hence
by at least $\gap(w)$. This bounds every $|u_c-u'_c|$ by $W_1/\gap(w)$ and recovers the labels. Second, near the
boundary of the simplex a single-linkage clustering of the atoms separates the class of a small coordinate
from the largest class, so each cluster holds at most $K-1$ classes and the local exponent improves to
$1/(2K-3)$, by the separated-cluster argument of \citet{wu2020optimal}. This absorbs the logarithmic
singularity of the entropy, so the transfer is uniform up to the boundary without a logarithmic factor.

\begin{proposition}[Sharpness]
  \label{prop:sharp}
  The odd moment system $X_1=X_3=\dots=X_{2K-3}=0$ has a nonzero real solution. For every such solution $x$
  and $0<s<\mu/\max_c|x_c|$, the reflected pair $\theta^\pm:=a\odot(\mu\one\pm sx)$ lies in $\Simplex$, has
  $\epsilon=2s^m|X_m(x)|$, and $\Ent(\theta^+)-\Ent(\theta^-)=2s\ell(x)+O(s^3)$ as $s\to0$. If $\gap(w)>0$, every such
  solution has $X_m(x)\ne0$; hence, if also $\ell(x)\ne0$, the exponent $1/m$ in \Cref{thm:modulus} cannot be
  improved.
\end{proposition}

\Cref{fig:modulus} in \Cref{app:modulus} shows the reflected family for $w=(1,2,4)$.

\begin{corollary}[Projection and Computation]
  \label{cor:projection}
  Let $\gap(w)>0$. For any $\hS$ and any $\tilde\theta\in\Simplex$ with $\max_v|S_v(\tilde\theta)-\hS_v|\le\min_\vartheta\max_v|S_v(\vartheta)-\hS_v|+\eta$,
  \[
    |\Ent(\tilde\theta)-\Ent(\theta)|\le C_K(w)\big(2\max_v|\hS_v-S_v(\theta)|+\eta\big)^{1/m}.
  \]
  On the grid $\{\vartheta\in\Simplex:M\vartheta\in\mathbb Z^K\}$ the minimizer is such a $\tilde\theta$ with
  $\eta=2(K-1)mU_1^{m-1}/M$, $U_1:=\max\{1,U\}$. With $1/M\asymp h:=\min\{1,\max\{1/n,1/N_m\}^{1/2}\}$ the grid
  version of \Cref{alg:main} keeps the bound of \Cref{thm:main}(i) and costs $O(n^2m+n^{(K-1)/2}Km)$ operations.
\end{corollary}

The corollary is proved in \Cref{app:modulus}.

\begin{proof}[Proof of \Cref{thm:main}(i)]
  By \Cref{cor:projection} with $\eta=0$, squaring, Jensen's inequality and \Cref{lem:var},
  \[
    \Exp(\hH-\Ent(\theta))^2\le4^{1/m}C_K(w)^2\Big(\sum_{v\le m}\Var(\hS_v)\Big)^{1/m}\lesssim\max\{1/n,1/\tilde N_m\}^{1/m},
  \]
  and $\tilde N_m\ge(4/5)^{m-1}N_m$; the risk is also at most $(\log K)^2$. Details are in \Cref{app:upper}.
\end{proof}

\section{The Lower Bound}
\label{sec:lower}

The lower bound compares two points through a path on which the channel moves.

\begin{figure}[t]
  \centering
  \includegraphics{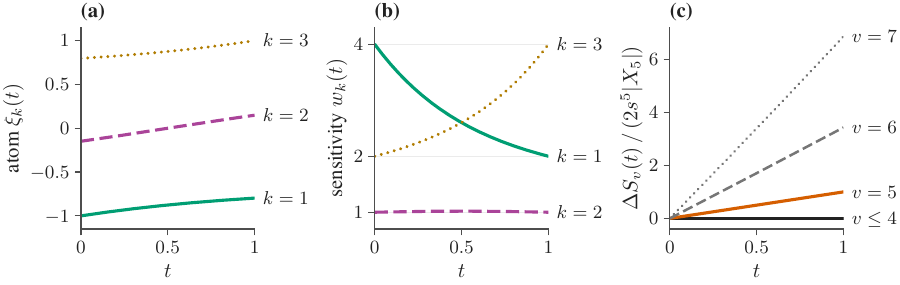}
  \caption{The path between the reflected pair $\theta^\pm=a\odot(\mu\one\pm sx)$ for $K=3$ and $w=(1,2,4)$
  (\Cref{lem:path}), where $a=1/w$, $\mu=1/\|a\|_1$, $x$ is the witness of \Cref{fig:simplex}(b) and
  $X_j(x)=\sum_ca_cx_c^j$; computed with $40$-digit arithmetic on a
  $201$-point grid. (a) Atoms $\xi_k(t)$ of the $3$-atomic measure whose moments of order at most four are
  those of $\nu_x=\sum_ca_c\delta_{x_c}$ and whose fifth moment is $(1-2t)X_5(x)$. (b) Sensitivities
  $w_k(t)=1/\alpha_k(t)$. Both endpoints carry the sensitivities $\{1,2,4\}$, attached to different atoms;
  in between, the channel leaves the known model, with computed values in $[1.000,4.000]$.
  (c) $\Delta S_v(t)/(2s^5|X_5(x)|)$, $\Delta S_v(t):=S_v(t)-S_v(0)$, at $s=0.05$: the weighted tree moments of order at most four do not
  move, $S_5(t)-S_5(0)=2t\,s^5|X_5(x)|$, and $S_6$ and $S_7$ move at the same order $s^5$.}
  \label{fig:path}
\end{figure}

\subsection{Two Points}
Let $x$ be a witness, normalized so that $\max_c|x_c|=1$. For small $s>0$ the reflected pair
$\theta^\pm=a\odot(\mu\one\pm sx)$ lies in $\Simplex$ because $X_1(x)=0$. Their locations are $u_c^\pm=\mu\pm sx_c$, so
$S_v(\theta^\pm)=\sum_j\binom vj\mu^{v-j}(\pm s)^jX_j(x)$: the even orders agree and the odd orders below $m$ vanish,
hence $S_1,\dots,S_{2K-2}$ coincide, while the entropies differ by $2s\ell(x)+O(s^3)$
(\Cref{prop:sharp}). If the laws of the verdicts under $\theta^+$ and $\theta^-$ are within total variation $1/2$
for $s\asymp\min\{1,N_m^{-1/(2m)}\}$, the two-point method gives $R^*_n(w)\gtrsim s^2\asymp\min\{1,N_m^{-1/m}\}$. We bound the
total variation by the Fisher information along a path from one law to the other:
$\TV(P_0,P_1)\le\frac12(\int_0^1I(t)\,\mathrm dt)^{1/2}$ (\Cref{lem:a:hellinger}).

\subsection{A Path on Which the Channel Moves}
With $w$ fixed, the map $(u_1,\dots,u_K)\mapsto(S_1,\dots,S_K)$ has Jacobian determinant
$K!\prod_ca_c\prod_{c<d}(u_d-u_c)$, which is nonzero at $\theta^\pm$, so no nonconstant path through $\theta^+$ keeps
even $S_1,\dots,S_K$ fixed (\Cref{lem:d:jacobian}). A path of distributions in which $S_1,\dots,S_{2K-2}$ stay
fixed must therefore leave the model with known $w$. The Hellinger bound needs only a smooth path of laws of
the verdicts, so this is allowed as long as both endpoints are the target laws.

\begin{lemma}[Moment-Preserving Path]
  \label{lem:path}
  Let $x$ be a witness and $\nu_x:=\sum_ca_c\delta_{x_c}$. For every $t\in\Real$ there is a unique positive measure
  $\nu_t=\sum_{k=1}^K\alpha_k(t)\delta_{\xi_k(t)}$ with at most $K$ atoms whose moments of orders $0,\dots,m-1$ are those of
  $\nu_x$ and whose moment of order $m$ is $(1-2t)X_m(x)$; the $\alpha_k,\xi_k$ are real-analytic, $\nu_0=\nu_x$ and
  $\nu_1=\nu_{-x}$. For $s$ small enough, the parameters
  \[
    \theta_k(t):=\alpha_k(t)\big(\mu+s\xi_k(t)\big),\qquad w_k(t):=1/\alpha_k(t),\qquad t\in[0,1],
  \]
  keep $S_1,\dots,S_{2K-2}$ fixed and ${\|\dot\theta\|_1+\|\dot w\|_\infty}=O(1)$, and for every $r$, on motifs with at
  most $r$ vertices, move every tree moment $S_v$, $v\ge m$, at rate $O(s^m)$, every unicyclic moment
  $\sum_k\theta_k^vw_k^v$ at rate $O(s)$ and every other motif moment $\sum_k\theta_k^vw_k^e$ at rate $O(1)$, with
  constants depending on $(K,w,x,r)$; at $t=0$ and $t=1$ they are $\theta^+$ and $\theta^-$ with the sensitivities $w$,
  up to relabeling the classes.
\end{lemma}

The positive definite Hankel matrix of $\nu_x$ forces the orthogonal polynomial with the prescribed top
moment to have $K$ simple real roots, which are the atoms of a Gauss-quadrature measure \citep{curto1991recursiveness,schmudgen2017moment}. We use this classical fact to build an analytic
interpolation with prescribed endpoints and with the derivative bounds that the information inequality
needs (\Cref{app:path}). \Cref{fig:path} shows the path for $w=(1,2,4)$.

\subsection{The Information Along the Path}
The verdict graph is a block model with diagonal edge probabilities $q+\Delta w_c$, and its law is a sum over
labelings. Following the homogeneous analyses \citep{anonymous2026four,anonymous2026three,anonymous2026rarity}, we start from a locally averaged score, a
function of labels and verdicts, and apply a sweep: a finite-order inclusion--exclusion over conditional
expectations given the verdicts and all but a few labels. The conditional expectation of the sweep given
the verdicts is the observed score, so by conditional Jensen, $I(t)$ is at most the second moment of the
sweep (\Cref{lem:a:sweep}).

We expand the sweep in the edge variables $\eps_{ij}:=\Delta(A_{ij}-q)/(q(1-q))$,
with $g_{ij}:=w_{Z_i}\Ind{Z_i=Z_j}$. Three features are new.
First, a verdict inside class $c$ has likelihood ratio exactly $1+w_c\eps_{ij}$, so the complete-data weight is
multilinear in $\eps$ and the score gains a channel term $\Ind{Z_i=Z_j}\dot w_{Z_i}\eps_{ij}/(1+w_{Z_i}\eps_{ij})$ because the
sensitivities move. Second, for motifs on at most $r$ vertices the Taylor coefficient of the sweep equals
$\partial_t[\eps^M]\log\Exp\prod_e(1+g_e\eps_e)$, a polynomial in the motif moments of the subgraphs of the motif
(\Cref{lem:e:exact}). By \Cref{lem:path}, trees contribute at order $s^m$ and unicyclic motifs at order $s$, and
Cauchy estimates bound every other coefficient by $O(1)$ (\Cref{lem:e:uniform}). Third, pairing overlapping
motifs splits by their cycle structure: trees give $N_ms^{2m}$, unicyclic motifs $\delta^3s^2$ (a cycle needs three
edges), multicyclic motifs $\delta^5/n$, and motifs on more than $r$ vertices $n\delta^r$. A bound on the remainder of the sweep after its Taylor polynomial of degree $D$, by a
radial expansion along $\lambda\eps$ on an event where all fields are small, completes the inequality
(\Cref{lem:e:info}): for $D\ge r+2$, $n$ large and $\delta\le1$ with $\sqrt\delta\log n$ small,
\[
  I(t)\le C\Big[N_ms^{2m}+\delta^3s^2+\frac{\delta^5}n+n\delta^r+n^{r+3}\delta^{D+1}(\log n)^{2(D-r)}+n^{4D+1}e^{-q_0^2\log^2n/36}\Big].
\]

\begin{proof}[Proof of \Cref{thm:main}(ii)]
  Given $\kappa$, take $r=\lfloor1/\kappa\rfloor+1$, so that $n\delta^r\le n^{1-\kappa r}\to0$, and $D\ge r+2$ with
  $\kappa(D+1)\ge r+4+\kappa$, so that the first remainder vanishes; the last term decays faster than every power
  of $n$, and the terms $\delta^3s^2$ and $\delta^5/n$ vanish on the window. With $s=\min\{s_1,a_0N_m^{-1/(2m)}\}$ for constants $s_1,a_0$, the tree term is at most $1/8$, so
  $\sup_tI(t)\le1/4$ for $n\ge n_L$ and $\TV(P_0,P_1)\le1/4$. The two-point bound (\Cref{lem:a:lecam}) and the
  entropy gap $s|\ell(x)|$ give $R^*_n(w)\ge c_L\min\{1,N_m^{-1/m}\}$ (\Cref{app:lower}).
\end{proof}

\section{Related Work}
\label{sec:related}

Four lines of work are closest to ours.

\subsection{Semantic Entropy and Weak Judges}
Semantic entropy groups sampled answers into meanings with a bidirectional entailment judge \citep{kuhn2023semantic}, and its discrete variant takes the entropy of the cluster proportions \citep{farquhar2024detecting}. Other uncertainty measures are built directly
from pairwise semantic similarities \citep{nikitin2024kernel,nguyen2025beyond}, whereas we estimate the
Shannon entropy of the latent meanings through a noisy binary channel. For a homogeneous pairwise model, a
spectral-clustering plug-in is consistent where clustering errors are small \citep{liu2026statistically}.

Minimax rates for a homogeneous judge are established for at most three meanings on the
whole weak window \citep[Cor.~3]{anonymous2026three}, for four meanings \citep[Thms.~1--2]{anonymous2026four}
and, graded by how many meanings can be rare, for general $K$ on polynomial windows \citep[Thm.~1, Cor.~1]{anonymous2026rarity}; there the least-favorable distributions approach a vertex of the
simplex. We share with those analyses the projection of unbiased moment
estimates and the sweep framework for the information. The meaning-dependent channel changes the
problem: the least-favorable family moves to the interior, the matched moment order doubles, the path
must move the channel, and the information inequality has to allow a moving channel.

\subsection{Moment Methods for Mixtures}
The unlabeled measure $\nu_\theta$ is a $K$-atomic mixing distribution with known masses. For $K$-atomic
distributions, moments of orders up to $2K-1$ control the Wasserstein distance with exponent $1/(2K-1)$,
separated clusters improve the exponent, and projecting noisy moments onto the moment space gives optimal
estimators of Gaussian mixtures \citep{wu2020optimal}; the moment comparison uses a sign-preserving polynomial in
the spirit of \citet{moitra2010settling}. Rates governed by the number of colliding atoms also appear for
over-fitted mixtures \citep{heinrich2018strong}, and moment matching drives the lower bounds for mixtures
of two Gaussians \citep{hardt2015tight}. Our additions are the class assignment through the subset-sum gap of
the known masses, the entropy transfer that is uniform up to the boundary without a logarithmic loss, the
genericity of witnesses for fixed masses, and an information bound for dependent pairwise observations
rather than independent samples.

\subsection{Truncated Moment Problem}
A moment sequence of orders up to $2K-2$ with a positive definite Hankel matrix, extended by any value of
order $2K-1$, has a unique $K$-atomic representing measure, the Gauss-quadrature measure of the associated
orthogonal polynomials \citep{curto1991recursiveness,schmudgen2017moment}. We use this classical fact to build
an analytic path with prescribed endpoints and bounded motif-moment derivatives.

\subsection{Block Models and Entropy Estimation}
The verdict graph is a stochastic block model with within-block probabilities $q+\Delta w_c$ and a common
between-block probability. Work on such models usually targets the partition, as in exact recovery \citep{abbe2016exact}, or the connectivity parameters. In the sparse two-community planted partition
model with edge probabilities $a/n$ and $b/n$, the parameters $a,b$ cannot be estimated consistently when
$(a-b)^2<2(a+b)$ \citep{mossel2015reconstruction}. In the dense regime the likelihood estimators are
asymptotically normal \citep{bickel2013asymptotic}, and, uniformly over models with nearly equal blocks, the
connectivity estimates converge slowly \citep{castillo2022uniform}. We fix the connectivity profile and
estimate the entropy of the unknown proportions as the signal vanishes. Our statistics are noise-corrected
subgraph averages \citep{chang2022estimation}, in the spirit of unbiased moment estimates along graph
structures \citep{kong2017spectrum}. With observed labels, entropy estimation on a fixed alphabet has rate $1/n$, and its
large-alphabet theory rests on polynomial approximation \citep{jiao2015minimax,wu2016minimax}.

\section{Conclusion}
\label{sec:conclusion}

A judge whose sensitivity depends on the meaning changes the minimax rate of semantic entropy even when
every sensitivity is known. For almost every sensitivity profile and $q$ bounded away from $0$ and $1$, the
risk is of order $\min\{1,(n^{2K-1}\Delta^{4K-4})^{-1/(2K-1)}\}$ on every fixed polynomial window, and the
consistency threshold, which needs no window, is that of a homogeneous judge on $2K-1$ meanings. The rate is attained by star statistics, a projection and a global weighted entropy modulus; the
matching lower bound comes from a moment-preserving path along which the channel moves. A judge with the
sensitivities $(1,2,4)$, which recognizes every meaning at least as well as the homogeneous judge with the same $q$
and $\Delta$, has the threshold $\beta<5/8$ for three meanings instead of $\beta<3/4$: how the sensitivity varies across
meanings, not its level, decides the rate. The difficulty moves with this variation, from rare meanings near a vertex of the
simplex to the interior point $\theta\propto1/w$, where every meaning is common.

\subsection{Limitations}
The matching lower bound holds on every fixed polynomial window $n\Delta^2\le n^{-\kappa}$, with constants that
depend on $\kappa$; on the whole weak window $n\Delta^2\le c_1$ it is not established, as for the homogeneous
analyses of general $K$. Neither the consistency threshold of \Cref{cor:threshold} nor the exponents in \Cref{fig:hero} depend on this restriction. The sensitivity profile, like $q$ and $\Delta$, is assumed known, and \Cref{alg:main} uses it. The lower bound,
proved with the profile known, holds a fortiori when it is unknown.

The constants depend on $w$ and are not uniform in it: the upper constant contains $1/\gap(w)$, and the lower constant contains $\ell(x)^2$ for the chosen witness. Both $\gap(w)$ and $\ell(x)$ tend to zero as $w$ approaches a homogeneous profile, so the fixed-profile results do not describe the transition between the homogeneous and the meaning-dependent rates. For a profile with tied sensitivities the upper bound is not established, although the lower bound of
\Cref{thm:main}(ii) covers such a profile whenever \Cref{cond:W} holds (\Cref{sec:results}). The constants $c_L$ and $n_L$ are not evaluated. The number of meanings $K$ is fixed, and the constants of
\Cref{thm:main} depend on it. Verdicts
are conditionally independent given the meanings, and the probability $q$ of a false equivalence is the same
for every pair of distinct meanings; both properties are used in the proof of the unbiasedness of the star statistics
(\Cref{lem:var}) and in the likelihood expansion of the lower bound (\Cref{sec:lower}).

\subsection{Outlook}
With $w$ unknown, cyclic motifs carry moments $\sum_c\theta_c^vw_c^v$ that tree statistics ignore, which suggests a
route to joint estimation of $\theta$ and $w$.

\section*{Acknowledgment}
This work was supported by the Young Doctoral Talent Incubation Program of the Second Affiliated Hospital, Army Medical
University, under Grant 2025YQB034. The authors declare no competing interests.

\bibliographystyle{plainnat}
\bibliography{references}

\clearpage
\appendix
\phantomsection\addcontentsline{toc}{section}{Appendix}%
\noindent{\Large\bfseries Appendix}\par
\section{Notation and General Tools}
\label{app:tools}

This appendix fixes the notation and proves the general tools of the lower bound.

\subsection{Notation}
$[K]=\{1,\dots,K\}$, $K\ge3$, $\Simplex=\{\theta\in[0,1]^K:\sum_c\theta_c=1\}$, and $m=2K-1$. The labels
$Z_1,\dots,Z_n$ are i.i.d.\ from $\theta$; given $Z$, the verdicts $A_{ij}$, $i<j$, are independent with
$A_{ij}\mid Z\sim\Bern(q+\Delta g_{ij})$, $g_{ij}:=w_{Z_i}\Ind{Z_i=Z_j}$. We write $a_c=1/w_c$, $\mu=1/\|a\|_1$,
$u_c=w_c\theta_c$, $\delta=n\Delta^2$, $N_v=n\delta^{v-1}$, $S_v(\theta)=\sum_c\theta_c^vw_c^{v-1}$ and
$m_{v,e}(\theta)=\sum_c\theta_c^vw_c^e$. The entropy uses the natural logarithm and $0\log0:=0$. Letters $C,c,L$
denote positive constants whose dependence is stated in each result. Estimators may use auxiliary
randomness independent of $(Z,A)$.

\subsection{Hellinger and Two-Point Bounds}
The lower bound uses three general tools. They hold for any smooth family of distributions on a finite
set and do not involve the channel; the framework of sweeps follows the homogeneous analyses \citep{anonymous2026four,anonymous2026three,anonymous2026rarity}. We state and prove them so that the paper
is self-contained.

\begin{lemma}[Hellinger Path Bound]
  \label{lem:a:hellinger}
  Let $\{P_t\}_{t\in[0,1]}$ be positive probability distributions on a finite set $\mathcal A$, continuously
  differentiable in $t$, and $I(t):=\sum_{A\in\mathcal A}P_t(A)(\partial_t\log P_t(A))^2$. Then
  $\TV(P_0,P_1)\le\frac12\big(\int_0^1I(t)\,\mathrm dt\big)^{1/2}$.
\end{lemma}

\begin{proof}
  Let $\mathcal H(P,Q):=\big(\sum_A(\sqrt{P(A)}-\sqrt{Q(A)})^2\big)^{1/2}$. By Cauchy--Schwarz,
  $\sum_A|P(A)-Q(A)|\le\mathcal H(P,Q)\big(\sum_A(\sqrt{P(A)}+\sqrt{Q(A)})^2\big)^{1/2}\le2\mathcal H(P,Q)$, so
  $\TV(P,Q)=\frac12\sum_A|P(A)-Q(A)|\le\mathcal H(P,Q)$. The map $t\mapsto(\sqrt{P_t(A)})_A\in\Real^{\mathcal A}$ is
  continuously differentiable with $\partial_t\sqrt{P_t(A)}=\frac12\sqrt{P_t(A)}\,\partial_t\log P_t(A)$, so
  \[
    \mathcal H(P_0,P_1)\le\int_0^1\Big(\sum_A\big(\partial_t\sqrt{P_t(A)}\big)^2\Big)^{1/2}\mathrm dt
    =\frac12\int_0^1\sqrt{I(t)}\,\mathrm dt\le\frac12\Big(\int_0^1I(t)\,\mathrm dt\Big)^{1/2}.\qedhere
  \]
\end{proof}

\begin{lemma}[Two-Point Bound]
  \label{lem:a:lecam}
  Let $\omega\ge0$. If $\theta^{(0)},\theta^{(1)}\in\Simplex$ satisfy $|H(\theta^{(0)})-H(\theta^{(1)})|\ge\omega$ and the laws of the
  observations under them satisfy $\TV(P_{\theta^{(0)}},P_{\theta^{(1)}})\le1/2$, then
  $R^*_n(w)\ge\omega^2/16$.
\end{lemma}

\begin{proof}
  For any estimator $\hH$, let $\mathcal E_i:=\{|\hH-H(\theta^{(i)})|\ge\omega/2\}$; since the two entropies are at least
  $\omega$ apart, $\mathcal E_0^c\subseteq\mathcal E_1$. Hence
  $P_{\theta^{(0)}}(\mathcal E_0)+P_{\theta^{(1)}}(\mathcal E_1)\ge P_{\theta^{(0)}}(\mathcal E_0)+P_{\theta^{(1)}}(\mathcal E_0^c)\ge1-\TV\ge1/2$ (the
  auxiliary randomness enters both laws identically). One of the two probabilities is at least $1/4$, and
  by Markov's inequality the corresponding risk is at least $(\omega/2)^2/4=\omega^2/16$ \citep[Sec.~2.3]{tsybakov2009introduction}.
\end{proof}

\subsection{Sweeps}
Fix $t$ and a parameter $(\theta(t),w(t))$ with $\theta_c(t)>0$, let $S_{\mathrm G}$ be a function of $(Z,A)$ (the
locally averaged score of \Cref{lem:e:glauber}) and $\Sobs:=\Exp_t[S_{\mathrm G}\mid A]$. For $U\subseteq[n]$ put
$\Pu_Uf:=\Exp_t[f\mid A,Z_{-U}]$, for $S\subseteq[n]$ put $\DS_S:=\sum_{U\subseteq S}(-1)^{|S|-|U|}\Pu_U$, and for an
integer $r\ge0$ let the sweep of order $r$ be $\Sw_r:=\sum_{|S|\le r}\DS_SS_{\mathrm G}$.

\begin{lemma}[Sweep]
  \label{lem:a:sweep}
  For $S\ne\emptyset$, $\Exp_t[\DS_Sf\mid A]=0$. Consequently $\Exp_t[\Sw_r\mid A]=\Sobs$, $\Exp_t\Sobs^2\le\Exp_t\Sw_r^2$,
  and $\sum_{S\subseteq[n]}\DS_S=\Pu_{[n]}$.
\end{lemma}

\begin{proof}
  $\sigma(A)\subseteq\sigma(A,Z_{-U})$, so $\Exp_t[\Pu_Uf\mid A]=\Exp_t[f\mid A]$ for every $U$ by the tower property.
  Hence $\Exp_t[\DS_Sf\mid A]=\Exp_t[f\mid A]\sum_{U\subseteq S}(-1)^{|S|-|U|}=0$ for $S\ne\emptyset$. Since
  $\DS_\emptyset=\Pu_\emptyset$ is the identity on functions of $(Z,A)$, we have $\Exp_t[\Sw_r\mid A]=\Exp_t[S_{\mathrm G}\mid A]=\Sobs$, and
  conditional Jensen gives $\Exp_t\Sobs^2\le\Exp_t\Sw_r^2$. Möbius inversion on the subsets of $[n]$ gives
  $\sum_{S\subseteq T}\DS_S=\Pu_T$ for every $T$, in particular for $T=[n]$.
\end{proof}

In the application, $\Sobs$ is the score of the observations, so $I(t)=\Exp_t\Sobs^2\le\Exp_t\Sw_r^2$
(\Cref{lem:e:glauber}).

\section{The Weighted Entropy Modulus}
\label{app:modulus}

This appendix proves \Cref{thm:modulus}, \Cref{prop:sharp} and \Cref{cor:projection}. Throughout,
$K\ge3$, $m=2K-1$, $w\in(0,\infty)^K$ with $\gap:=\gap(w)>0$, $U=\max_cw_c$, $w_{\min}:=\min_cw_c$,
$a_{\max}:=1/w_{\min}=\max_ca_c$, and $u_c=w_c\theta_c$, $u'_c=w_c\theta'_c$ for $\theta,\theta'\in\Simplex$.

\subsection{The Moment Measure}

For $\theta\in\Simplex$ the measure $\nu_\theta=\sum_ca_c\delta_{u_c}$ is a positive measure on $[0,U]$ of
total mass $\|a\|_1$, and $\int x^v\,\mathrm d\nu_\theta=S_v(\theta)$ for $v\ge0$, with $S_0=\|a\|_1$ and
$S_1=\sum_c\theta_c=1$. Fix $\theta,\theta'$, put $\epsilon:=\max_{2\le v\le m}|S_v(\theta)-S_v(\theta')|$ and
\[
  G(x):=\nu_\theta((-\infty,x])-\nu_{\theta'}((-\infty,x]),\qquad W:=\int_{\Real}|G(x)|\,\mathrm dx ,
\]
so that $W$ is the Wasserstein-1 distance between $\nu_\theta$ and $\nu_{\theta'}$. Let $z_1<\dots<z_\ell$
($\ell\le2K$) be the distinct atoms of the two measures. $G$ is right-continuous, equal to a constant $g_i$
on $[z_i,z_{i+1})$, zero on $(-\infty,z_1)$ and on $[z_\ell,\infty)$ because the two masses agree, and
$|G|\le\|a\|_1$. If $\ell\le1$ then $G\equiv0$, so let $\ell\ge2$. With $I(x):=\{c:u_c\le x\}$ and
$I'(x):=\{c:u'_c\le x\}$,
\begin{equation}
  \label{eq:b:G}
  G(x)=\sum_{c\in I(x)\setminus I'(x)}a_c-\sum_{c\in I'(x)\setminus I(x)}a_c .
\end{equation}

\subsection{Four Lemmas}

The first two lemmas follow the moment-comparison argument of \citet[Prop.~7]{wu2020optimal},
which uses a sign-preserving polynomial in the spirit of \citet{moitra2010settling}; here the measures
have total mass $\|a\|_1$, support in $[0,U]$ and the common first moment $S_1=1$, and the constants are
explicit. \Cref{lem:b:boundary} is the counterpart of their Proposition~4 for separated clusters.

\begin{lemma}[Sign-Preserving Polynomial]
  \label{lem:b:sign}
  If $G\equiv0$, let $P\equiv1$ and $r=0$. Otherwise let the nonzero values among the $g_i$ occur on the
  intervals with indices $i_1<\dots<i_p$, $p\ge1$, and for each $s<p$ with $g_{i_s}g_{i_{s+1}}<0$ choose a
  point in the closed interval $[z_{i_s+1},z_{i_{s+1}}]$. The chosen points $t_1<\dots<t_r$ satisfy
  $r\le\ell-2\le2K-2$ and $t_k\in[0,U]$, and for some $\sigma\in\{\pm1\}$ the polynomial
  $P(x):=\sigma\prod_k(x-t_k)$ satisfies $P(x)G(x)\ge0$ for all $x\in\Real$.
\end{lemma}

\begin{proof}
  $r\le p-1\le\ell-2$. The closed intervals $[z_{i_s+1},z_{i_{s+1}}]$ for different $s$ are disjoint, and
  none meets the interior $(z_{i_t},z_{i_t+1})$ of an interval where $G\ne0$. Hence $P$ has a constant sign
  on the interior of each such interval, and between $i_s$ and $i_{s+1}$ it changes sign exactly when
  $g_{i_s}$ and $g_{i_{s+1}}$ have opposite signs. Choosing $\sigma$ so that $P$ has the sign of $g_{i_1}$ on
  the first such interval gives $PG\ge0$ on the interiors; by continuity of $P$ the inequality extends to
  their left endpoints, and $G=0$ elsewhere.
\end{proof}

\begin{lemma}[Moment Comparison]
  \label{lem:b:moments}
  If $\epsilon\le1$, then $W\le C_1\epsilon^{1/m}$ with $C_1:=(1+U)^{2K-2}/2+{(4K-4)\|a\|_1}$.
\end{lemma}

\begin{proof}
  Take $P$ from \Cref{lem:b:sign} and $Q(x):=\int_0^xP$, a polynomial of degree $r+1\le m$. By Fubini and
  $(\nu_\theta-\nu_{\theta'})(\Real)=0$,
  \[
    \int P(x)G(x)\,\mathrm dx=\int\Big(\int_y^UP(x)\,\mathrm dx\Big)\,\mathrm d(\nu_\theta-\nu_{\theta'})(y)
    =-\int Q\,\mathrm d(\nu_\theta-\nu_{\theta'}).
  \]
  Writing $P(x)=\sum_{j=0}^rp_jx^j$, we get $\int Q\,\mathrm d(\nu_\theta-\nu_{\theta'})=
  \sum_{j=0}^rp_j(S_{j+1}(\theta)-S_{j+1}(\theta'))/(j+1)$. The term $j=0$ vanishes because $S_1\equiv1$,
  and the others involve orders between $2$ and $m$. Since $|t_k|\le U$,
  we have $\sum_j|p_j|\le\prod_k(1+|t_k|)\le(1+U)^r$, so
  \begin{equation}
    \label{eq:b:PG}
    \int|P||G|\,\mathrm dx=\int PG\,\mathrm dx\le\tfrac12(1+U)^r\epsilon .
  \end{equation}
  If $r=0$ then $P\equiv\sigma$, only the term $j=0$ remains, and $W=\int PG=0$. Let $r\ge1$. Since
  $|P(x)|=\prod_k|x-t_k|$ exceeds $\eta$ outside $\bigcup_k[t_k-\eta^{1/r},t_k+\eta^{1/r}]$, the set
  $\{|P|\le\eta\}$ has Lebesgue measure at most $2r\eta^{1/r}$. Hence for every $\eta>0$,
  \[
    W\le\eta^{-1}\int|P||G|+\|a\|_1\,\mathrm{Leb}\{|P|\le\eta\}\le\frac{(1+U)^r\epsilon}{2\eta}+2r\|a\|_1\eta^{1/r}.
  \]
  If $\epsilon=0$, letting $\eta\downarrow0$ gives $W=0$. Otherwise $\eta=\epsilon^{r/(r+1)}$ gives
  $W\le[(1+U)^r/2+2r\|a\|_1]\epsilon^{1/(r+1)}$, and $r+1\le m$, $\epsilon\le1$ give the claim.
\end{proof}

\begin{lemma}[Class Assignment]
  \label{lem:b:assign}
  $\max_c|u_c-u'_c|\le W/\gap$.
\end{lemma}

\begin{proof}
  If $I(x)\ne I'(x)$, the sets $I(x)\setminus I'(x)$ and $I'(x)\setminus I(x)$ are disjoint and not both
  empty, so $|G(x)|\ge\gap$ by \eqref{eq:b:G}. For every class $c$ and every $x$ in the half-open interval
  between $\min(u_c,u'_c)$ and $\max(u_c,u'_c)$, exactly one of $u_c\le x$ and $u'_c\le x$ holds, so $c$ lies
  in the symmetric difference of $I(x)$ and $I'(x)$ and $I(x)\ne I'(x)$. Hence
  $|u_c-u'_c|\le\mathrm{Leb}\{x:I(x)\ne I'(x)\}\le\gap^{-1}\int|G|=W/\gap$.
\end{proof}

\Cref{lem:b:moments,lem:b:assign} give \Cref{thm:modulus}(a). The gap $\gap>0$ enters only through
\Cref{lem:b:assign}, which turns closeness of the unlabeled measures into closeness of the labeled
parameters; colliding atoms ($u_c=u_{c'}$) and atoms at $0$ ($\theta_c=0$) affect neither lemma.

\begin{lemma}[Near the Boundary]
  \label{lem:b:boundary}
  Let $\rho_b:=w_{\min}/(4KU)$, $d_0:=w_{\min}/(8K^2)$, $\tau:=w_{\min}/(4K^2)$, $b_0:=\min\{1,\tau/2\}$,
  $\epsilon_1:=b_0^{(2K-2)(2K-3)/(2K-4)}$ and
  $C_2:=(1+U)^{2K-2}/2+\lfloor K/2\rfloor(4K-8)\|a\|_1b_0^{-(2K-2)/(2K-4)}$. If
  $\min_c\min(\theta_c,\theta'_c)<\rho_b$, $D:=\max_c|u_c-u'_c|<d_0$ and $\epsilon<\epsilon_1$, then
  $W\le C_2\epsilon^{1/(2K-3)}$.
\end{lemma}

\begin{proof}
  If $\epsilon=0$, \Cref{lem:b:moments} gives $W=0$; let $\epsilon>0$. By symmetry let $\theta_k<\rho_b$, and
  let $\theta_L=\max_c\theta_c\ge1/K$, so $k\ne L$. Write $J_c$ for the closed interval between $u_c$ and
  $u'_c$, of length at most $D<d_0$. The points of $J_L$ exceed $w_{\min}/K-d_0$ and those of $J_k$ are below
  $U\rho_b+d_0=w_{\min}/(4K)+d_0$, so for $K\ge3$
  \[
    \mathrm{dist}(J_k,J_L)>\tfrac{3w_{\min}}{4K}-2d_0\ge\tfrac{2w_{\min}}{3K}.
  \]

  \emph{Single-linkage clusters.} Join $J_c$ and $J_{c'}$ when $\mathrm{dist}(J_c,J_{c'})<\tau$ and take
  connected components. If $J_k$ and $J_L$ were in one component, a chain $J_k=Q_0,\dots,Q_s=J_L$ with
  $s\le K-1$ and consecutive distances below $\tau$ would give $\mathrm{dist}(J_k,J_L)<(K-1)\tau+(K-2)d_0<
  3w_{\min}/(8K)$, a contradiction. So every component has at most $K-1$ classes.

  \emph{The convex hulls of different components are at distance at least $\tau$.} Let $\mathcal C\ne\mathcal C'$
  be components with convex hulls $\mathcal H,\mathcal H'$. If the hulls are disjoint and closer than $\tau$,
  the two facing endpoints belong to intervals of $\mathcal C$ and $\mathcal C'$ at distance below $\tau$,
  which would have been joined. If the hulls meet, one of them, say $\mathcal H$, contains an endpoint $y$
  of an interval $J_{c'}$ with $c'\in\mathcal C'$. The intervals of $\mathcal C$, together with the connecting
  segments of length below $\tau$, form a connected subset of $\Real$ that contains both endpoints of
  $\mathcal H$ and hence covers $\mathcal H$; so $y$ lies on an interval of $\mathcal C$ or on a segment of length
  below $\tau$, and in both cases, $J_{c'}$ is at distance below $\tau$ from an interval of $\mathcal C$, a
  contradiction. Consequently $u_c$ and $u'_c$ lie in the same hull for every $c$, and $I(x)=I'(x)$, so
  $G(x)=0$, on the gaps between hulls.

  \emph{Location of the roots.} Use the freedom in \Cref{lem:b:sign}: a sign change between nonzero
  intervals in different components gets its root at the midpoint of a gap between them (the closed
  interval of \Cref{lem:b:sign} covers that gap), and a sign change inside one component has its root in
  that component's hull. A component with $s_{\mathcal C}$ classes has at most $2s_{\mathcal C}$ jump points in its
  hull and at most $2s_{\mathcal C}-2\le2K-4$ interior roots; a one-class component has none. For $x$ in a
  hull $\mathcal H_{\mathcal C}$, every root outside $\mathcal H_{\mathcal C}$ is at distance at least
  $\tau/2\ge b_0$ from $x$, and $b_0\le1$ and there are at most $2K-2$ roots in total, so
  $|P(x)|\ge b_0^{2K-2}\prod_{t_k\in\mathcal H_{\mathcal C}}|x-t_k|$.

  \emph{Estimate.} Take $\eta:=\epsilon^{(2K-4)/(2K-3)}$; then $\epsilon<\epsilon_1$ is equivalent to
  $\lambda:=\eta b_0^{-(2K-2)}<1$. On a component without interior roots, $|P|\ge b_0^{2K-2}>\eta$. On a
  component with $r_{\mathcal C}\ge1$ interior roots, $\{x\in\mathcal H_{\mathcal C}:|P(x)|\le\eta\}\subseteq
  \{x:\prod_{t_k\in\mathcal H_{\mathcal C}}|x-t_k|\le\lambda\}$, of measure at most $2r_{\mathcal C}\lambda^{1/r_{\mathcal C}}
  \le(4K-8)\lambda^{1/(2K-4)}$. There are at most $\lfloor K/2\rfloor$ such components and $G=0$ outside
  the hulls, so by \eqref{eq:b:PG}
  \begin{align*}
    W&\le\eta^{-1}\int|P||G|+\|a\|_1\,\mathrm{Leb}(\{|P|\le\eta\}\cap\{G\ne0\})\\
    &\le\frac{(1+U)^{2K-2}\epsilon}{2\eta}+\Big\lfloor\frac K2\Big\rfloor(4K-8)\|a\|_1b_0^{-\frac{2K-2}{2K-4}}\eta^{\frac1{2K-4}}
    =C_2\epsilon^{\frac1{2K-3}}.\qedhere
  \end{align*}
\end{proof}

\subsection{Proof of \texorpdfstring{\Cref{thm:modulus}}{Theorem 2}(b)}

Let $h(p):=-p\log p$. We use three elementary facts.
\begin{enumerate}[label=(h\arabic*),leftmargin=*]
  \item If $p,p'\in[\rho_b,1]$, then $|h(p)-h(p')|\le(1+\log(1/\rho_b))|p-p'|$, since $|h'(x)|=|1+\log x|$.
  \item If $p,p'\in[0,1]$ and $t:=|p-p'|\le1/2$, then $|h(p)-h(p')|\le h(t)$. Indeed $h$ is concave, so
    $x\mapsto h(x+t)-h(x)$ is nonincreasing on $[0,1-t]$ with values in $[-h(1-t),h(t)]$, and
    $g(t):=h(t)-h(1-t)$ is concave on $[0,1/2]$ ($g''(t)=1/(1-t)-1/t\le0$) with $g(0)=g(1/2)=0$, so
    $h(1-t)\le h(t)$.
  \item For $x\in(0,1]$, $x\log(1/x)\le\frac m{2e}x^{(2K-3)/m}$, because $x^{2/m}\log(1/x)\le m/(2e)$.
\end{enumerate}
Write $\Theta:=\max_c|\theta_c-\theta'_c|\le a_{\max}D$ with $D=\max_c|u_c-u'_c|$, and note
$\|\theta-\theta'\|_1=\sum_ca_c|u_c-u'_c|\le\|a\|_1D$. We bound $|\Delta H|:=|H(\theta)-H(\theta')|$ by a constant
times $\epsilon^{1/m}$ case by case, with $\rho_b,d_0,\epsilon_1,C_2$ from \Cref{lem:b:boundary}.
\begin{enumerate}[label=\arabic*.,leftmargin=*,start=0]
  \item If $\epsilon=0$, \Cref{lem:b:moments,lem:b:assign} give $\theta=\theta'$. If $\epsilon>1$,
    $|\Delta H|\le\log K\le\log K\cdot\epsilon^{1/m}$. Let $0<\epsilon\le1$, so that
    $D\le C_1\gap^{-1}\epsilon^{1/m}$ by part (a).
  \item Interior, all $\theta_c,\theta'_c\ge\rho_b$: by (h1),
    $|\Delta H|\le(1+\log(1/\rho_b))\|\theta-\theta'\|_1\le(1+\log(1/\rho_b))\|a\|_1C_1\gap^{-1}\epsilon^{1/m}$.
  \item Boundary (some $\theta_c$ or $\theta'_c$ below $\rho_b$) and $D\ge d_0$: then
    $\epsilon^{1/m}\ge\gap d_0/C_1$, so $|\Delta H|\le\log K\cdot C_1(\gap d_0)^{-1}\epsilon^{1/m}$.
  \item Boundary, $D<d_0$, $\epsilon\ge\epsilon_1$: $|\Delta H|\le\log K\cdot\epsilon_1^{-1/m}\epsilon^{1/m}$.
  \item Boundary, $D<d_0$, $\epsilon<\epsilon_1$: by \Cref{lem:b:assign,lem:b:boundary},
    $\Theta\le a_{\max}W/\gap\le a_{\max}C_2\gap^{-1}\epsilon^{1/(2K-3)}$. If $\Theta\le1/e$, every
    $|\theta_c-\theta'_c|\le1/2$ and $h$ increases on $[0,1/e]$, so by (h2) and (h3),
    $|\Delta H|\le\sum_ch(|\theta_c-\theta'_c|)\le Kh(\Theta)\le\frac{Km}{2e}\Theta^{(2K-3)/m}\le
    \frac{Km}{2e}(a_{\max}C_2/\gap)^{(2K-3)/m}\epsilon^{1/m}$. If $\Theta>1/e$, then
    $\epsilon^{1/(2K-3)}>\gap/(ea_{\max}C_2)$ and $|\Delta H|\le\log K<\log K\cdot(ea_{\max}C_2/\gap)^{(2K-3)/m}\epsilon^{1/m}$.
\end{enumerate}
Hence \Cref{thm:modulus}(b) holds with
\begin{equation}
  \label{eq:b:CK}
  \begin{aligned}
  C_K(w):=\max\Big\{&\frac{(1+\log(1/\rho_b))\|a\|_1C_1}{\gap},\ \frac{Km}{2e}\Big(\frac{a_{\max}C_2}{\gap}\Big)^{\frac{2K-3}m},\\
  &\log K\cdot\max\Big\{1,\frac{C_1}{\gap d_0},\epsilon_1^{-\frac1m},\Big(\frac{ea_{\max}C_2}{\gap}\Big)^{\frac{2K-3}m}\Big\}\Big\},
  \end{aligned}
\end{equation}
which depends only on $K$, $U$, $w_{\min}$, $\|a\|_1$ and $\gap$. \qed

\subsection{Proof of \texorpdfstring{\Cref{prop:sharp}}{Proposition 2}}

The map $x\mapsto(X_1(x),X_3(x),\dots,X_{2K-3}(x))$ from the unit sphere $S^{K-1}$ to $\Real^{K-1}$ is
continuous and odd, so by the Borsuk--Ulam theorem \citep[Thm.~2.1.1]{matousek2008using} it has a zero
$x$. The measures $\nu_x:=\sum_ca_c\delta_{x_c}$ and $\nu_{-x}$ have the same moments of orders $0$ to $m-1$:
the even ones by symmetry and the odd ones because they vanish. If also $X_m(x)=0$, all moments of orders
$0,\dots,m$ agree; the signed measure $\nu_x-\nu_{-x}$ has at most $2K=m+1$ atoms, so the Vandermonde system
forces $\nu_x=\nu_{-x}$. The proof of \Cref{lem:b:assign} uses only \eqref{eq:b:G}, which holds for any two
measures $\sum_ca_c\delta_{y_c}$ and $\sum_ca_c\delta_{y'_c}$ with $y,y'\in\Real^K$; applied to $y=x$ and $y'=-x$ it gives
$|x_c-(-x_c)|=0$ for every $c$, that is, $x=0$. Hence $X_m(x)\ne0$ when $\gap>0$; this is the only part of the
proposition that uses $\gap>0$.

For $0<s<\mu/\max_c|x_c|$, $\theta^\pm:=a\odot(\mu\one\pm sx)$ lie in $\Simplex$, since
$\sum_ca_c(\mu\pm sx_c)=\mu\|a\|_1\pm sX_1(x)=1$. Since $u^\pm_c=\mu\pm sx_c$, the binomial theorem gives
$S_v(\theta^\pm)=\sum_{j=0}^v\binom vj\mu^{v-j}(\pm s)^jX_j(x)$, so
\[
  S_v(\theta^+)-S_v(\theta^-)=2\sum_{j\ \mathrm{odd}}\binom vj\mu^{v-j}s^jX_j(x),
\]
which vanishes for $v<m$ and equals $2s^mX_m(x)$ for $v=m$; thus $\epsilon=2s^m|X_m(x)|$. On the entropy
side, $f(s):=H(a\odot(\mu\one+sx))$ is analytic for $|s|<\mu/\max_c|x_c|$, $H(\theta^+)-H(\theta^-)=f(s)-f(-s)$
is odd in $s$, and $f'(0)=-\sum_ca_cx_c(\log(a_c\mu)+1)=\ell(x)-(\log\mu+1)X_1(x)=\ell(x)$. Hence
$H(\theta^+)-H(\theta^-)=2s\ell(x)+O(s^3)$ and
\[
  \frac{|H(\theta^+)-H(\theta^-)|}{\epsilon^{1/m}}\ \longrightarrow\ \frac{2^{1-1/m}|\ell(x)|}{|X_m(x)|^{1/m}}\qquad(s\to0),
\]
which is positive when $\ell(x)\ne0$. Therefore no bound $|\Delta H|\le C\epsilon^{\kappa'}$ with
$\kappa'>1/m$ holds near $\theta^\star=\mu a$. \qed

\begin{figure}[!t]
  \centering
  \includegraphics{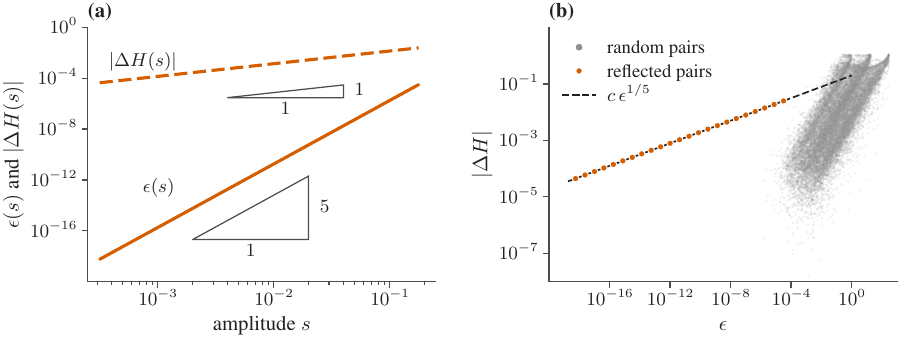}
  \caption{Reflected pairs attain the exponent $1/5$ of \Cref{thm:modulus} ($K=3$, $w=(1,2,4)$, $a=1/w$,
  $\mu=1/\|a\|_1$, and the witness $x$ of \Cref{fig:simplex}(b), for which $\ell(x)=-\sum_ca_cx_c\log a_c\approx-0.0703$).
  (a) Along $\theta^\pm=a\odot(\mu\one\pm sx)$, computed with $40$-digit arithmetic,
  $\epsilon(s)=\max_{2\le v\le5}|S_v(\theta^+)-S_v(\theta^-)|=2|X_5(x)|\,s^5$ and $|\Delta H(s)|/s\to2|\ell(x)|$.
  (b) $|\Delta H|$ against $\epsilon$ for the reflected pairs and for $19{,}950$ random pairs in double
  precision: $\theta$ is Dirichlet with all parameters equal to $0.3$, $1$ or $5$, and $\theta'$ adds a Gaussian
  perturbation of scale between $10^{-4}$ and $1$, clipped at zero and renormalized; pairs with
  $\epsilon\le10^{-12}$ are dropped. The dashed line $c\,\epsilon^{1/5}$ with $c\approx0.1985$ is the asymptote of the
  reflected family, not a bound for the random pairs.}
  \label{fig:modulus}
\end{figure}

\subsection{Proof of \texorpdfstring{\Cref{cor:projection}}{Corollary 2}}

Let $F(\vartheta):=\max_{2\le v\le m}|S_v(\vartheta)-\hS_v|$, continuous on the compact set $\Simplex$, so an
exact minimizer $\htheta$ exists. For any $\tilde\theta\in\Simplex$ with $F(\tilde\theta)\le F(\htheta)+\eta$,
\[
  \epsilon(\tilde\theta,\theta)\le F(\tilde\theta)+F(\theta)\le F(\htheta)+\eta+F(\theta)\le2F(\theta)+\eta,
\]
and \Cref{thm:modulus}(b) gives $|H(\tilde\theta)-H(\theta)|\le C_K(w)(2\max_v|\hS_v-S_v(\theta)|+\eta)^{1/m}$;
$\eta=0$ is the statement for $\htheta$. Squaring, using $(x+y)^{2/m}\le x^{2/m}+y^{2/m}$ (as $2/m<1$), Jensen's
inequality for the concave map $x\mapsto x^{1/m}$ and $\max_v|\cdot|^2\le\sum_v|\cdot|^2$,
\begin{equation}
  \label{eq:b:risk}
  \Exp(H(\tilde\theta)-H(\theta))^2\le C_K(w)^2\Big[4^{1/m}\Big(\sum_{v=2}^m\Exp(\hS_v-S_v(\theta))^2\Big)^{1/m}+\eta^{2/m}\Big].
\end{equation}

\emph{Grid.} For an integer $M\ge1$ let $\mathcal G_M:=\{\vartheta\in\Simplex:M\vartheta\in\mathbb Z^K\}$, which has
$\binom{M+K-1}{K-1}=O(M^{K-1})$ points. Every $\vartheta\in\Simplex$ is within $\ell_1$-distance $2(K-1)/M$ of
$\mathcal G_M$: round the first $K-1$ coordinates down to multiples of $1/M$ and give the remaining mass to the
last coordinate. For $\vartheta,\vartheta'\in\Simplex$ and $v\le m$,
$|S_v(\vartheta)-S_v(\vartheta')|\le\sum_cw_c^{v-1}|\vartheta_c^v-\vartheta_c'^v|\le vU_1^{v-1}\|\vartheta-\vartheta'\|_1$, so
$F$ is $mU_1^{m-1}$-Lipschitz in $\ell_1$. A minimizer $\tilde\theta$ of $F$ over $\mathcal G_M$ therefore satisfies
$F(\tilde\theta)\le F(\htheta)+\eta$ with $\eta:=2(K-1)mU_1^{m-1}/M$. With
$1/M\asymp h:=\min\{1,\max\{1/n,1/N_m\}^{1/2}\}$, which equals $\min\{1,N_m^{-1/2}\}$ when $n\Delta^2\le1$,
we have $\eta^{2/m}\lesssim\max\{1/n,1/N_m\}^{1/m}$, so \eqref{eq:b:risk} and \Cref{lem:var} give the rate of
\Cref{thm:main}(i) for $\tilde\theta$. The grid has $O(h^{-(K-1)})=O(n^{(K-1)/2})$ points and evaluating $F$ at
one point costs $O(Km)$ operations, so with \Cref{lem:var} the whole estimator costs
$O(n^2m+n^{(K-1)/2}Km)$ operations. \qed

\section{Moment Statistics and the Upper Bound}
\label{app:upper}

This appendix proves \Cref{lem:var} and \Cref{thm:main}(i). Write $X_{ij}:=(A_{ij}-q)/\Delta$, so that
$\Exp[X_{ij}\mid Z]=g_{ij}$ and $|g_{ij}|\le U$. Recall
$p_c=q+\Delta w_c$, $\rho=\max\{q(1-q),\max_cp_c(1-p_c)\}/\Delta^2$, $\tilde N_v=n^v/(U^2+\rho)^{v-1}$ and
$U_1=\max\{1,U\}$; $\rho$ bounds the conditional variance of one verdict divided by $\Delta^2$.

Let $T$ be a tree on the vertex set $\{1,\dots,v\}$, let $\mathcal I$ be the set of injective maps
$\varphi:\{1,\dots,v\}\to[n]$, $|\mathcal I|=(n)_v$, and put $Y_\varphi:=\prod_{\{a,b\}\in E(T)}X_{\varphi(a)\varphi(b)}$ and
$\hS_T:=|\mathcal I|^{-1}\sum_{\varphi\in\mathcal I}Y_\varphi$. When $T$ is the star with center $1$, each star on $v$
answers with a given center is produced by $(v-1)!$ maps, so $\hS_T$ is the statistic $\hS_v$ of
\Cref{alg:main}:
\[
  \hS_v=\frac{(v-1)!}{(n)_v}\sum_{i=1}^ne_{v-1}\big(X_{ij}:j\ne i\big).
\]
For each $i$, the recursion $e_k\leftarrow e_k+X_{ij}e_{k-1}$, run over $j\ne i$ with $k$ decreasing from
$m-1$ to $1$ and $e_0=1$, computes $e_1,\dots,e_{m-1}$ in $O(nm)$ operations, so all of
$\hS_2,\dots,\hS_m$ cost $O(n^2m)$ operations.

\begin{lemma}[Restatement of \Cref{lem:var}, for Every Tree]
  \label{lem:c:var}
  For every tree $T$ on $v\ge2$ vertices, $\Exp\hS_T=S_v(\theta)$, and for $n\ge2v-2$
  \[
    \Var(\hS_T)\le2^{v+1}\sum_{s=1}^v\binom vs(v)_s\,\frac{U^{2(v-s)}(U^2+\rho)^{s-1}}{n^s}
    \le B_v\max\{1/n,1/\tilde N_v\},
  \]
  where $B_v:=2^{2v+1}v!\,U_1^{2v-2}$.
\end{lemma}

\begin{proof}
  \emph{Unbiasedness.} Given $Z$, distinct pairs are independent, and the $v-1$ edges of $\varphi(T)$ are
  distinct, so $\Exp[Y_\varphi\mid Z]=\prod_{\{a,b\}\in E(T)}g_{\varphi(a)\varphi(b)}$. Since $T$ is connected,
  this equals $w_c^{v-1}$ if all $Z_{\varphi(a)}$ equal a common $c$ and $0$ otherwise; taking expectations
  gives $\sum_c\theta_c^vw_c^{v-1}=S_v(\theta)$.

  \emph{Covariances.} $\Var(\hS_T)=|\mathcal I|^{-2}\sum_{\varphi,\varphi'}\Cov(Y_\varphi,Y_{\varphi'})$. The variable
  $Y_\varphi$ depends only on the labels in $\mathrm{im}\,\varphi$ and the verdicts inside $\mathrm{im}\,\varphi$. If
  $\mathrm{im}\,\varphi\cap\mathrm{im}\,\varphi'=\emptyset$ these two families are independent (labels are
  independent, and given the labels, the verdicts are independent and depend only on their endpoints), so
  $\Cov(Y_\varphi,Y_{\varphi'})=0$. Let $|\mathrm{im}\,\varphi\cap\mathrm{im}\,\varphi'|=s\ge1$ and let the two embedded
  trees share $t$ edges. The shared edges form a subforest of $\varphi(T)$ with all endpoints in the
  intersection; if $t\ge1$ this forest has at least $t+1$ vertices, so $t\le s-1$ (trivially also when
  $t=0$). Given $Z$,
  \[
    \Exp[Y_\varphi Y_{\varphi'}\mid Z]=\prod_{e\ \text{in one tree only}}g_e\cdot\prod_{e\ \text{shared}}\Exp[X_e^2\mid Z],
  \]
  with $|g_e|\le U$ and $\Exp[X_e^2\mid Z]=g_e^2+\Var(A_e\mid Z)/\Delta^2\le U^2+\rho$. There are
  $2(v-1)-2t$ unshared edges, so
  $|\Exp[Y_\varphi Y_{\varphi'}]|\le U^{2(v-1-t)}(U^2+\rho)^t\le U^{2(v-s)}(U^2+\rho)^{s-1}$, the last step because the
  middle expression is nondecreasing in $t$ and $t\le s-1$. Also $S_v\le U^{v-1}$, so $S_v^2$ obeys the same
  bound and $|\Cov(Y_\varphi,Y_{\varphi'})|\le2U^{2(v-s)}(U^2+\rho)^{s-1}$.

  \emph{Counting.} The number of ordered pairs $(\varphi,\varphi')$ with intersection size $s$ is
  $(n)_v\binom vs(v)_s(n-v)_{v-s}$: choose which $s$ positions of $\varphi'$ land in $\mathrm{im}\,\varphi$, map
  them injectively into $\mathrm{im}\,\varphi$, and map the rest injectively outside it. For $n\ge2v-2$,
  $(n)_v\ge(n/2)^v$, so the fraction of such pairs is at most $\binom vs(v)_s2^vn^{-s}$. Multiplying and
  summing over $s$ gives the first inequality.

  \emph{Simplification.} Put $\lambda:=(U^2+\rho)/n$. Then $U^{2(v-s)}\le U_1^{2v-2}$,
  $(U^2+\rho)^{s-1}/n^s=\lambda^{s-1}/n\le\max\{1,\lambda\}^{v-1}/n$ and $\sum_s\binom vs(v)_s\le2^vv!$, so
  $\Var(\hS_T)\le2^{2v+1}v!\,U_1^{2v-2}\max\{1,\lambda^{v-1}\}/n$, and $\lambda^{v-1}/n=1/\tilde N_v$.
\end{proof}

\begin{proof}[Proof of \Cref{thm:main}(i)]
  $H(\htheta)$ and $H(\theta)$ lie in $[0,\log K]$, so the risk is at most $(\log K)^2$. By
  \eqref{eq:b:risk} with $\eta=0$ and \Cref{lem:c:var},
  \[
    \Exp(H(\htheta)-H(\theta))^2\le4^{1/m}C_K(w)^2\Big(\sum_{v=2}^m\Var(\hS_v)\Big)^{1/m}.
  \]
  For $2\le v\le m$, $B_v\le B_m$ and $\max\{1/n,1/\tilde N_v\}\le\max\{1/n,1/\tilde N_m\}$: with
  $\lambda=(U^2+\rho)/n$, if $\lambda\ge1$ then $1/\tilde N_v=\lambda^{v-1}/n\le\lambda^{m-1}/n$, and if $\lambda<1$
  then $1/\tilde N_v\le1/n$. Hence $\sum_v\Var(\hS_v)\le(m-1)B_m\max\{1/n,1/\tilde N_m\}$ and
  \[
    \begin{aligned}
    \Exp(H(\htheta)-H(\theta))^2\le\min\Big\{&(\log K)^2,\\
    &4^{1/m}C_K(w)^2\big((m-1)B_m\big)^{1/m}\max\{1/n,1/\tilde N_m\}^{1/m}\Big\}.
    \end{aligned}
  \]
  This holds for every $q\ge0$ and $\Delta>0$ with $q+\Delta U\le1$. In that range, $U\Delta\le1$ and
  $\rho\le1/(4\Delta^2)$, so $\Delta^2(U^2+\rho)\le U^2\Delta^2+1/4\le5/4$ and
  $\tilde N_m\ge(4/5)^{m-1}N_m$; therefore $\max\{1/n,1/\tilde N_m\}\le(5/4)^{m-1}\max\{1/n,1/N_m\}$, which gives
  the stated bound with $C_U:=(5/4)^{(m-1)/m}4^{1/m}C_K(w)^2((m-1)B_m)^{1/m}$. If $n\Delta^2\le1$, then
  $N_m=n(n\Delta^2)^{m-1}\le n$ and the maximum is $1/N_m$. Finally, $(\vartheta,\hS)\mapsto\max_v|S_v(\vartheta)-\hS_v|$
  is continuous and $\Simplex$ is compact, so a measurable minimizer exists by the measurable maximum
  theorem.
\end{proof}

\section{The Moment-Preserving Path}
\label{app:path}

This appendix proves \Cref{lem:path} and the statement in \Cref{sec:lower} that the channel must move.
Fix $w$ satisfying \Cref{cond:W} with a witness $x$, normalized so that $\max_c|x_c|=1$, and write
$\nu_x:=\sum_ca_c\delta_{x_c}$ and $X_j:=X_j(x)=\int\xi^j\,\mathrm d\nu_x(\xi)$. For $t\in\Real$ let
$M_j(t):=X_j$ for $0\le j\le m-1$ and $M_m(t):=(1-2t)X_m$. Motif moments of a parameter $(\theta,w)$ are
$m_{v,e}:=\sum_k\theta_k^vw_k^e$; the tree moments are $S_v=m_{v,v-1}$ and the unicyclic ones $m_{v,v}$.

\begin{lemma}[Restatement of \Cref{lem:path}]
  \label{lem:d:path}
  \begin{enumerate}[label=(\roman*),leftmargin=*]
    \item For every $t\in\Real$ there is a unique positive measure with at most $K$ atoms,
      $\nu_t=\sum_{k=1}^K\alpha_k(t)\delta_{\xi_k(t)}$ with $\xi_1<\dots<\xi_K$ and $\alpha_k>0$, whose
      moments of orders $0,\dots,m$ are $M_0,\dots,M_m(t)$. The functions $\alpha_k,\xi_k$ are real-analytic
      on $\Real$, $\nu_0=\nu_x$ and $\nu_1=\nu_{-x}$.
    \item Let $\Xi:=\max_{t\in[0,1],k}|\xi_k(t)|$, let $\alpha_{\mathrm{lo}},\alpha_{\mathrm{hi}}$ be the minimum and
      maximum of the $\alpha_k(t)$ over $t\in[0,1]$, and $s_P:=\mu/(2\Xi)$. For $0<s\le s_P$ and
      $t\in[0,1]$ put
      \[
        \theta_k(t):=\alpha_k(t)\big(\mu+s\xi_k(t)\big),\qquad w_k(t):=1/\alpha_k(t).
      \]
      Then $\theta(t)\in\Simplex$, $\theta_k(t)\ge\alpha_{\mathrm{lo}}\mu/2$, $w_k(t)\in[w_{\mathrm{lo}},w_{\mathrm{hi}}]:=
      [1/\alpha_{\mathrm{hi}},1/\alpha_{\mathrm{lo}}]$, and $\theta,w$ are real-analytic on a neighborhood of $[0,1]$.
    \item $S_v(t)=\sum_{j=0}^v\binom vj\mu^{v-j}s^jM_j(t)$, where $M_j(t):=\int\xi^j\,\mathrm d\nu_t$; in particular
      $S_v(t)$ does not depend on $t$ for $1\le v\le m-1$.
    \item For every $r$ there is $L\ge1$ depending only on $(K,w,x,r)$ such that for $0<s\le\min\{s_P,1\}$,
      $t\in[0,1]$, $2\le v\le r$ and $v-1\le e\le\binom v2$,
      \[
        |\dot m_{v,v-1}|\le Ls^m,\qquad|\dot m_{v,v}|\le Ls,\qquad|\dot m_{v,e}|\le L\ \ (e\ge v+1),
      \]
      and $\eta(t):=\|\dot\theta(t)\|_1+\|\dot w(t)\|_\infty\le L$, $\max_k|\dot\theta_k|/\theta_k\le L$; dots
      denote $\partial_t$.
    \item As multisets of classes $\{(\theta_k,w_k)\}$, the parameter at $t=0$ is $\theta^+=a\odot(\mu\one+sx)$ with
      sensitivities $w$, and the parameter at $t=1$ is $\theta^-=a\odot(\mu\one-sx)$ with sensitivities $w$.
    \item There is $s_H>0$ depending only on $(w,x)$ such that $|H(\theta^+)-H(\theta^-)|\ge s|\ell(x)|$ for
      $0<s\le s_H$.
  \end{enumerate}
\end{lemma}

\begin{proof}
  (i) $\nu_x$ has $K$ distinct atoms and positive masses, so the Hankel matrix $\mathbf H:=[X_{i+j}]_{0\le i,j\le K-1}$
  is positive definite; it does not depend on $t$. Define the monic polynomial $P_t(z)=z^K+\sum_{j<K}b_jz^j$
  by $\mathbf Hb=-(M_K,\dots,M_m(t))^\top$, so that $b$ is affine in $t$. Let $\mathcal L_t(p):=\sum_jp_jM_j(t)$ for
  polynomials of degree at most $m$; then $\mathcal L_t(P_tz^i)=0$ for $i<K$. If $P_t$ had fewer than $K$ real roots
  of odd multiplicity, let $Q$ be the product of the corresponding linear factors, $\deg Q<K$; then $P_tQ$ is a
  nonzero nonnegative polynomial of degree at most $2K-2$, hence a sum of two squares of polynomials of
  degree at most $K-1$, and $\mathcal L_t(P_tQ)>0$ by positive definiteness of $\mathbf H$, contradicting
  orthogonality. So $P_t$ has $K$ distinct simple real roots $\xi_k(t)$. With the Lagrange basis $\ell_k$
  (degree $K-1$) put $\alpha_k:=\mathcal L_t(\ell_k)$; since $\ell_k^2-\ell_k$ vanishes at all roots, it equals $P_t$
  times a polynomial of degree at most $K-2$, so $\alpha_k=\mathcal L_t(\ell_k^2)>0$. Every polynomial $p$ of degree
  at most $m$ is $P_tr$ plus its interpolation polynomial on the roots ($\deg r\le K-1$), so
  $\mathcal L_t(p)=\sum_k\alpha_kp(\xi_k)$: $\nu_t$ has the required moments.

  Uniqueness: a representing positive
  measure with fewer than $K$ atoms would give $\mathcal L_t(Q^2)=0$ for the product $Q$ of its atom factors,
  contradicting positive definiteness; one with exactly $K$ atoms has a monic atom polynomial solving the
  same linear system as $P_t$, hence equal to $P_t$, and its masses are fixed by the Vandermonde system. The
  coefficients of $P_t$ are affine in $t$ and its roots are simple, so the $\xi_k$ are real-analytic by the
  implicit function theorem, and so are the $\alpha_k$, given by the Vandermonde system. The moments of
  $\nu_{-x}$ are $(-1)^jX_j$: they equal $X_j$ for even $j$, vanish for odd $j\le m-2$, and equal $-X_m$ for
  $j=m$, which are the $M_j(1)$; uniqueness gives $\nu_0=\nu_x$ and $\nu_1=\nu_{-x}$ (as the classical theory
  of the truncated moment problem predicts; see \citealp{curto1991recursiveness,schmudgen2017moment}).

  (ii) $\sum_k\theta_k=\mu\sum_k\alpha_k+s\sum_k\alpha_k\xi_k=\mu M_0+sM_1=1$, since $M_0=\|a\|_1$ and
  $M_1=X_1=0$. For $s\le s_P$, $\mu+s\xi_k\ge\mu/2$. The rest follows from continuity and compactness.

  (iii) $\theta_k^vw_k^{v-1}=\alpha_k(\mu+s\xi_k)^v$; expand binomially. For $j\le m-1$, $M_j(t)$ does not
  depend on $t$.

  (iv) In general $m_{v,e}=\sum_k\alpha_k^{v-e}(\mu+s\xi_k)^v$. For $e=v-1$, by (iii),
  $\dot m_{v,v-1}=\sum_{j=m}^v\binom vj\mu^{v-j}s^j\dot M_j(t)$, and the $\dot M_j$ are bounded on $[0,1]$ and $s\le1$,
  so $|\dot m_{v,v-1}|\le Ls^m$ (and $\dot m_{v,v-1}=0$ for $v<m$). For $e=v$, $m_{v,v}=\sum_k(\mu+s\xi_k)^v$ and
  $\dot m_{v,v}=s\sum_kv(\mu+s\xi_k)^{v-1}\dot\xi_k$. For $e\ge v+1$ differentiate directly. Also
  $\dot\theta_k=\dot\alpha_k(\mu+s\xi_k)+s\alpha_k\dot\xi_k$, $\dot w_k=-\dot\alpha_k/\alpha_k^2$ and
  $\dot\theta_k/\theta_k=\dot\alpha_k/\alpha_k+s\dot\xi_k/(\mu+s\xi_k)$. Every quantity is controlled by the bounds of
  $\alpha,\dot\alpha,1/\alpha,\xi,\dot\xi$ on $[0,1]$ and by $\mu$, uniformly in $s$.

  (v) At $t=0$, $\{(\alpha_k,\xi_k)\}=\{(a_c,x_c)\}$, so $\{(\theta_k,w_k)\}=\{(a_c(\mu+sx_c),w_c)\}$; at $t=1$ the
  same holds with $x$ replaced by $-x$. The law of the verdict graph depends only on the multiset of
  classes $\{(\theta_c,w_c)\}$.

  (vi) $f(\sigma):=H(a\odot(\mu\one+\sigma x))$ is analytic for $|\sigma|<\mu$; $H(\theta^+)-H(\theta^-)=f(s)-f(-s)$
  is odd in $s$ and equals $2sf'(0)+O(s^3)$, with $f'(0)=\ell(x)$ as in the proof of \Cref{prop:sharp}.
\end{proof}

\begin{lemma}[The Channel Must Move]
  \label{lem:d:jacobian}
  For fixed masses $a\in(0,\infty)^K$ and locations $u\in\Real^K$, the map $u\mapsto(S_1,\dots,S_K)$ with
  $S_j=\sum_ca_cu_c^j$ has Jacobian determinant
  \[
    \det\Big[\frac{\partial S_j}{\partial u_c}\Big]_{j,c=1}^K=K!\,\Big(\prod_ca_c\Big)\prod_{c<d}(u_d-u_c).
  \]
  Consequently, with $w$ fixed, no nonconstant continuous path $t\mapsto\theta(t)$ through an endpoint of the
  reflected pair $\theta^\pm$ of a witness (\Cref{prop:sharp}) keeps $S_1,\dots,S_K$ constant.
\end{lemma}

\begin{proof}
  $\partial S_j/\partial u_c=ja_cu_c^{j-1}$, so the matrix is $\mathrm{diag}(1,\dots,K)\,V\,\mathrm{diag}(a)$ with the
  Vandermonde matrix $V=[u_c^{j-1}]_{j,c}$, whose determinant is $\prod_{c<d}(u_d-u_c)$. At an endpoint of a
  reflected pair the locations $u_c=\mu\pm sx_c$ are distinct, so the determinant is nonzero and, since
  $\theta\mapsto u=w\odot\theta$ is a linear bijection for fixed $w$, the map $\theta\mapsto(S_1,\dots,S_K)$ is a
  local diffeomorphism there and in particular injective on a neighborhood of the endpoint. For a
  continuous path on an interval along which $S_1,\dots,S_K$ are constant, the set of times at which the path
  sits at the endpoint is closed by continuity and open by this local injectivity, so it is the whole
  interval.
\end{proof}

The path keeps $S_1,\dots,S_{2K-2}$ fixed and $2K-2\ge K$, so by \Cref{lem:d:jacobian} it cannot keep $w$
fixed. Because the Hellinger bound of \Cref{lem:a:hellinger} needs only a smooth path of distributions of the
observations, the intermediate models may leave the experiment with known $w$; both endpoints belong to it
by \Cref{lem:d:path}(v).

\section{The Information Inequality}
\label{app:information}

This appendix bounds the Fisher information of the verdict graph along the path of \Cref{lem:d:path}.
Fix such a path (for some $s\in(0,\min\{s_P,1\}]$) and $t\in[0,1]$, write $P_t$ for the law of the verdicts
under $(\theta(t),w(t))$, $\ell_c:=\dot\theta_c/\theta_c$, $\eta:=\|\dot\theta\|_1+\|\dot w\|_\infty$, and let $w_{\mathrm{hi}}$ and
$L$ be the constants of \Cref{lem:d:path}. We assume
\begin{equation}
  \tag{A}
  \label{eq:e:A}
  q\in[q_0,1-q_0],\qquad\Delta\le\Delta_0:=q_0/(8\max\{1,w_{\mathrm{hi}}\}).
\end{equation}

\subsection{Edge Variables, the Glauber Identity and the Score}

\begin{lemma}[Edge Variables]
  \label{lem:e:edge}
  Let $\eps_{ij}:=\Delta(A_{ij}-q)/(q(1-q))$ and $g_{ij}(Z;t):=w_{Z_i}(t)\Ind{Z_i=Z_j}$.
  \begin{enumerate}[label=(\roman*),leftmargin=*]
    \item $P_t(A\mid Z)=\prod_{i<j}\Bern(A_{ij};q)\cdot\prod_{i<j}(1+g_{ij}(Z;t)\eps_{ij})$.
    \item $|\eps_{ij}|\le2\Delta/q_0$, hence $|g_{ij}\eps_{ij}|\le1/4$.
    \item Given $Z$, the $\eps_{ij}$ are independent and $\Exp_t[\eps_{ij}\mid Z]=\Delta^2g_{ij}/(q(1-q))$. With
      $C_\eps:=4\max\{1,w_{\mathrm{hi}}\}/q_0$, so that $C_\eps\Delta\le1/2$, for every integer $a\ge1$,
      $|\Exp_t[\eps_{ij}^a\mid Z]|\le(C_\eps\Delta)^{c(a)}$, where $c(1):=2$ and $c(a):=a$ for $a\ge2$.
  \end{enumerate}
\end{lemma}

\begin{proof}
  (i) For $b\in\{0,1\}$, $\Bern(b;q+\Delta g)/\Bern(b;q)=1+\Delta g(b-q)/(q(1-q))$. (ii) $|A-q|\le1$ and
  $q(1-q)\ge q_0/2$. (iii) $\Exp[A_{ij}-q\mid Z]=\Delta g_{ij}$, so $|\Exp[\eps_{ij}\mid Z]|\le2w_{\mathrm{hi}}\Delta^2/q_0\le
  (C_\eps\Delta)^2$; for $a\ge2$ use (ii).
\end{proof}

The likelihood ratio of a verdict inside class $c$ is exactly $1+w_c\eps$, and the complete-data weight is
multilinear in the edge variables; this is what makes the expansions below exact.

\begin{lemma}[Glauber Identity and Score]
  \label{lem:e:glauber}
  Let $\theta_c>0$ for all $c$.
  \begin{enumerate}[label=(\roman*),leftmargin=*]
    \item For $U\subseteq[n]$, the conditional law of $Z_U$ given $(A,Z_{-U})$ is proportional to
      $z'_U\mapsto\prod_{k\in U}\theta_{z'_k}\prod_{e\cap U\ne\emptyset}(1+g_e(z)\eps_e)$, $z=(z'_U,Z_{-U})$. In particular
      $P_t(Z_i=c\mid A,Z_{-i})=\theta_cf_{ic}/\sum_b\theta_bf_{ib}$ with $f_{ic}:=\prod_{j\ne i:Z_j=c}(1+w_c\eps_{ij})$.
    \item The score of the observations $\Sobs:=\partial_t\log P_t(A)$ satisfies $\Sobs=\Exp_t[\SG\mid A]$ with
      \[
        \SG:=\sum_i\Phi(f_i)+\sum_{i<j}\chi_{ij},\qquad\Phi(f):=\frac{\sum_c\dot\theta_cf_c}{\sum_c\theta_cf_c},\qquad
        \chi_{ij}:=\Ind{Z_i=Z_j}\frac{\dot w_{Z_i}\eps_{ij}}{1+w_{Z_i}\eps_{ij}} .
      \]
      We call $\SG$ the locally averaged score: it replaces each label term $\ell_{Z_i}$ of the complete-data
      score $\sum_i\ell_{Z_i}+\sum_{i<j}\chi_{ij}$ by its conditional expectation given $(A,Z_{-i})$.
      Moreover $\Exp_t\Sobs=0$ and $I(t)=\Exp_t\Sobs^2$; at $\eps=0$, $f_i=\one$, $\Phi(f_i)=\sum_c\dot\theta_c=0$
      and $\chi=0$; for positive $f$, $|\Phi(f)|\le\max_c|\ell_c|$; and $|\chi_{ij}|\le2\|\dot w\|_\infty|\eps_{ij}|$.
  \end{enumerate}
\end{lemma}

\begin{proof}
  (i) By \Cref{lem:e:edge}(i), the Bernoulli factors and the factors of edges not meeting $U$ do not depend
  on $z'_U$. For $U=\{i\}$, $\prod_j(1+g_{ij}\eps_{ij})=\prod_{j:Z_j=z'_i}(1+w_{z'_i}\eps_{ij})=f_{i,z'_i}$.
  (ii) $P_t(A)=\sum_z\prod_i\theta_{z_i}\cdot\prod\Bern(A_e;q)\cdot\prod_e(1+g_e(z;t)\eps_e)$ is a finite sum.
  Differentiating term by term with $\partial_tg_{ij}=\dot w_{z_i}\Ind{z_i=z_j}$ gives
  $\Sobs=\Exp_t[\sum_i\ell_{Z_i}+\sum_{i<j}\chi_{ij}\mid A]$, and by (i) and the tower property,
  $\Exp_t[\ell_{Z_i}\mid A]=\Exp_t[\Phi(f_i)\mid A]$. Also $\Exp_t\Sobs=\sum_A\partial_tP_t(A)=0$. The bounds use
  \Cref{lem:e:edge}(ii).
\end{proof}

\subsection{The Sweep as an Analytic Function of the Edge Variables}
Fix labels $z\in[K]^n$ and let $\mathsf w(z;\eps):=\prod_v\theta_{z_v}\prod_{k<l}(1+g_{kl}(z)\eps_{kl})$ and
\[
  \Pu_Uf(z):=\sum_{z'_U}\mathsf w(z'_U,z_{-U})f(z'_U,z_{-U})\Big/\sum_{z'_U}\mathsf w(z'_U,z_{-U}),
\]
which is the conditional
expectation of \Cref{app:tools} when $\eps=\eps(A)$, by \Cref{lem:e:glauber}(i). Writing $f_{ic}$ as
$\prod_{j\ne i:z_j=c}(1+w_c\eps_{ij})$, $\SG$ and each $\DS_S\SG$ are rational functions of $\eps$; after the
factors depending only on the fixed $z_{-U}$ cancel, each conditional normalizer equals $1$ at $\eps=0$, so
$\Sw_r=\mathcal S_r(\eps;z)$ is analytic on a complex neighborhood of $\eps=0$. A \emph{motif} $M$ is a multiset
of pairs in $[n]$ with multiplicities $m_e$; $\eps^M:=\prod_e\eps_e^{m_e}$, $|M|:=\sum_em_e$, $V(M)$ is the set of
vertices it touches, $v(M):=|V(M)|$, $d(M)$ is the number of distinct pairs, $G_M$ is the underlying simple
graph and $\beta(G):=d-v+1$ is the cyclomatic number of a connected graph. Let
$B^{(r)}_M(z):=[\eps^M]\,\mathcal S_r(\cdot;z)$. The \emph{motif system} on a vertex set $V$ is the same
construction with only the pairs inside $V$; it is marked by a superscript $(V)$.

\begin{lemma}[Locality]
  \label{lem:e:local}
  Let $M$ be a motif and $V:=V(M)$. (i) If $S\not\subseteq V$, then $[\eps^M]\,\DS_S\SG=0$. (ii) If $S\subseteq V$,
  then $[\eps^M]\,\DS_S\SG(z)=[\eps^M]\,\DS^{(V)}_S\SG^{(V)}(z_V)$. (iii) Hence $B^{(r)}_M$ depends only on $z_V$ and
  $B^{(r)}_M=[\eps^M]\sum_{S\subseteq V,|S|\le r}\DS^{(V)}_S\SG^{(V)}$. (iv) If $M$ is not connected, $B^{(r)}_M=0$.
\end{lemma}

\begin{proof}
  The coefficient $[\eps^M]$ of an analytic function depends only on its restriction to the subspace
  $\{\eps_e=0,\ e\notin M\}$, on which we compute. A vertex $u\notin V$ has all its edge variables equal to $0$:
  its factor in $\mathsf w$ is $\theta_{z_u}$ alone, $f_u=\one$ and $\Phi(f_u)=0$; for every other vertex $i$, the
  factor $1+w_c\eps_{iu}$ of $f_{ic}$ equals $1$, so $f_i$ does not depend on $z_u$; and $\chi_{iu}=0$. Hence on
  the subspace, $\SG=\SG^{(V)}(z_V)$ and $\mathsf w(z;\eps)=\prod_{u\notin V}\theta_{z_u}\cdot\mathsf w^{(V)}(z_V;\eps)$, so the
  conditional law of $z'_U$ is the product of the $V$-system posterior of $z'_{U\cap V}$ and the prior of
  $z'_{U\setminus V}$. For $f$ depending on $z_V$ only, this gives $\Pu_Uf=\Pu^{(V)}_{U\cap V}f$.

  (i) Pick $u\in S\setminus V$. In $\DS_S=\sum_{U\subseteq S}(-1)^{|S|-|U|}\Pu_U$, pair $U$ with $U\triangle\{u\}$: the two
  terms carry opposite signs and $\Pu_U\SG=\Pu^{(V)}_{U\cap V}\SG^{(V)}=\Pu_{U\triangle\{u\}}\SG$, so they cancel.
  (ii) For $S\subseteq V$ every $U\subseteq S$ satisfies $U\cap V=U$. (iii) follows from (i) and (ii).
  (iv) Let $V=V_1\cup V_2$ with no pair of $M$ between $V_1$ and $V_2$. On the subspace the $V$-system is the
  product of the $V_1$- and $V_2$-systems: $\Phi(f_i)$ for $i\in V_1$ involves only pairs inside $V_1$, $\chi_{ij}=0$
  for $i\in V_1$, $j\in V_2$, and the posterior factorizes. So $\SG^{(V)}=\SG^{(V_1)}+\SG^{(V_2)}$. Applying the
  argument of (i) to $\SG^{(V_1)}$, with $V_1$ in place of $V$, gives $\DS^{(V)}_S\SG^{(V_1)}=0$ unless $S\subseteq V_1$,
  in which case $\DS^{(V)}_S\SG^{(V_1)}=\DS^{(V_1)}_S\SG^{(V_1)}$ involves no pair meeting $V_2$; the same holds with
  the indices exchanged. No term contains $\eps^M$, which uses pairs of both parts, so $B^{(r)}_M=0$.
\end{proof}

\begin{lemma}[Exact Identity]
  \label{lem:e:exact}
  For every motif $M$, $V:=V(M)$, $r\ge0$ and $z$,
  \[
    B^{(r)}_M(z)=\partial_tC_M-\sum_{S\subseteq V,|S|>r}[\eps^M]\big(\DS^{(V)}_S\SG^{(V)}\big)(z),\qquad
    C_M:=[\eps^M]\log\Exp_{\theta(t)}\prod_{e\subseteq V}\big(1+g_e(Z;t)\eps_e\big),
  \]
  with $Z\sim\theta(t)^{\otimes V}$. $C_M$ is a polynomial, with coefficients depending only on $M$, in the motif
  moments $m_{v(Q),e(Q)}(t)$ of the connected simple subgraphs $Q$ of $G_M$ ($e(Q)$ the number of edges). If
  $G_M$ is a tree, only tree moments $S_{v'}$, $2\le v'\le v(M)$, occur; if $\beta(G_M)=1$, only tree moments
  and unicyclic moments $m_{v',v'}$.
\end{lemma}

\begin{proof}
  On the motif system, $\sum_{S\subseteq V}\DS^{(V)}_S=\Pu^{(V)}_V$ (\Cref{lem:a:sweep}), so by \Cref{lem:e:local}(iii),
  $B^{(r)}_M=[\eps^M]\Pu^{(V)}_V\SG^{(V)}-\sum_{|S|>r}[\eps^M]\DS^{(V)}_S\SG^{(V)}$. As a function of $z_i$, $\mathsf w_V(z)$ is
  $\theta_{z_i}f_{i,z_i}$ times a factor free of $z_i$, and $f_i$ does not depend on $z_i$, so
  $\sum_{z_i}\mathsf w_V\ell_{z_i}=\Phi(f_i)\sum_{z_i}\mathsf w_V$. Moreover
  $\partial_t\mathsf w_V=\mathsf w_V\cdot\big(\sum_{i\in V}\ell_{z_i}+\sum_{e\subseteq V}\partial_tg_e\eps_e/(1+g_e\eps_e)\big)$, and the second sum is
  $\sum_e\chi_e$. Therefore
  \[
    \Pu^{(V)}_V\SG^{(V)}=\frac{\sum_z\mathsf w_V(z)\,\partial_t\log\mathsf w_V(z)}{\sum_z\mathsf w_V(z)}=\partial_t\log\sum_z\mathsf w_V(z)
    =\partial_t\log\Exp_\theta\prod_e(1+g_e\eps_e).
  \]
  The function $\log\sum_z\mathsf w_V$ is jointly analytic in $t$ near $[0,1]$ and in $\eps$ near $0$ (the path is analytic
  and $\sum_z\mathsf w_V=1$ at $\eps=0$), so $[\eps^M]$ and $\partial_t$ commute. For the structure, expand
  $\Exp\prod_e(1+g_e\eps_e)=\sum_{B\subseteq E_V}\eps^B\,\Exp\prod_{e\in B}g_e$. The product $\prod_{e\in B}g_e(Z)$ is nonzero only if
  every connected component $Q$ of $(V(B),B)$ is monochromatic, and then
  $\Exp\prod_{e\in B}g_e=\prod_Q\sum_c\theta_c^{v(Q)}w_c^{e(Q)}=\prod_Qm_{v(Q),e(Q)}$. Expanding $\log(1+y)$, the coefficient
  $[\eps^M]$ involves only $B\subseteq\mathrm{supp}\,M$, i.e., subgraphs of $G_M$, whose components are connected
  subgraphs of $G_M$: trees if $G_M$ is a tree, and trees or unicyclic graphs if $\beta(G_M)=1$.
\end{proof}

\subsection{Coefficients}

\begin{lemma}[Small Motifs]
  \label{lem:e:small}
  If $v(M)\le r$, then $B^{(r)}_M=\partial_tC_M$ does not depend on $z$, and the following hold: $B^{(r)}_M=0$ if $G_M$ is a tree
  and $v(M)\le m-1$; $|B^{(r)}_M|\le\Gamma_MLs^m$ if $G_M$ is a tree and $m\le v(M)\le r$; $|B^{(r)}_M|\le\Gamma_MLs$
  if $\beta(G_M)=1$; and $|B^{(r)}_M|\le\Gamma_ML$ if $\beta(G_M)\ge2$. Here $\Gamma_M$ depends only on $M$ and
  $w_{\mathrm{hi}}$.
\end{lemma}

\begin{proof}
  For $v(M)\le r$ the sum over $|S|>r$ in \Cref{lem:e:exact} is empty. $C_M$ is a polynomial $Q_M$ of the motif
  moments, which lie in $[0,\max\{1,w_{\mathrm{hi}}\}^{\binom r2}]$, where the partial derivatives of $Q_M$ are
  bounded; differentiate by the chain rule. For a tree, only tree moments occur: those with $v'\le m-1$
  have zero derivative and the others derivative $O(Ls^m)$ (\Cref{lem:d:path}(iii),(iv)); for $\beta=1$ the
  unicyclic moments add derivatives $O(Ls)$; in general every derivative is $O(L)$.
\end{proof}

\begin{lemma}[Uniform Coefficient Bound]
  \label{lem:e:uniform}
  For every motif $M$ with $v:=v(M)\ge2$, every $r\ge0$ and every $z\in[K]^V$,
  \[
    |B^{(r)}_M(z)|\le c_v\varrho_v^{-|M|}\eta,\qquad\varrho_v:=\frac{\log(3/2)}{\binom v2\max\{1,w_{\mathrm{hi}}\}},\qquad
    c_v:=3^{v+1}v^2\max\{1,1/w_{\mathrm{hi}}\}.
  \]
\end{lemma}

\begin{proof}
  On the motif system take the complex polydisc $\mathbb D:=\{|\eps_e|\le\varrho_v,\ e\subseteq V\}$. (1) Each product
  $\prod_e(1+g_e\eps_e)$ differs from $1$ by at most $\prod(1+w_{\mathrm{hi}}\varrho_v)-1\le e^{\binom v2w_{\mathrm{hi}}\varrho_v}-1\le1/2$,
  so every conditional normalizer $\sum_{z'_U}\prod\theta\prod(1+g_e\eps_e)$, which equals $1$ at $\eps=0$, has modulus at
  least $1/2$, and the sum of the moduli of its terms is at most $3/2$. (2) $|f_{ic}-1|\le e^{(v-1)w_{\mathrm{hi}}\varrho_v}-1\le1/2$,
  so $|\sum_c\theta_cf_{ic}|\ge1/2$; since $\sum_c\dot\theta_c=0$, we have $|\sum_c\dot\theta_cf_{ic}|=|\sum_c\dot\theta_c(f_{ic}-1)|\le\|\dot\theta\|_1/2$
  and $|\Phi(f_i)|\le\|\dot\theta\|_1$. (3) $w_{\mathrm{hi}}\varrho_v\le\log(3/2)<1/2$, so
  $|\chi_e|\le\|\dot w\|_\infty\varrho_v/(1-w_{\mathrm{hi}}\varrho_v)\le\|\dot w\|_\infty/w_{\mathrm{hi}}$. Hence
  $|\SG^{(V)}|\le v\|\dot\theta\|_1+\binom v2\|\dot w\|_\infty/w_{\mathrm{hi}}\le(c_v/3^{v+1})\eta$ on $\mathbb D$,
  $|\Pu^{(V)}_U\SG^{(V)}|\le3c_v\eta/3^{v+1}$, and $|\DS^{(V)}_S\SG^{(V)}|$ is at most $2^{|S|}$ times this. The Cauchy estimate
  $|[\eps^M]f|\le\sup_{\mathbb D}|f|/\varrho_v^{|M|}$ and $\sum_{S\subseteq V}2^{|S|}=3^v$ give the claim via \Cref{lem:e:local}(iii).
\end{proof}

\subsection{Pairing}

Let $\Pi_D:=\sum_{1\le|M|\le D}\eps^MB^{(r)}_M(Z)$, summed over all motifs on $[n]$. Since $\SG=0$ at $\eps=0$,
$\Pi_D$ is the Taylor polynomial of degree $D$ of $\Sw_r$ in $\eps$.

\begin{lemma}[Pairing]
  \label{lem:e:pairing}
  Assume \eqref{eq:e:A}, the conclusions of \Cref{lem:d:path} and $\delta\le1$. Then
  \[
    \Var_t(\Pi_D)\le C_D\big[N_ms^{2m}+\delta^3s^2+\delta^5/n+n\delta^r\big],
  \]
  with $C_D$ depending only on $(K,q_0,r,D,L,w_{\mathrm{hi}})$.
\end{lemma}

\begin{proof}
  Split the motifs into five classes: \emph{zero} (not connected, or $v(M)\le\min\{r,m-1\}$ with $G_M$ a tree;
  coefficient identically $0$ by \Cref{lem:e:local}(iv) and \Cref{lem:e:small}); \emph{tree} ($v(M)\le r$, $G_M$ a
  tree, $v(M)\ge m$); \emph{unicyclic} ($v(M)\le r$, $\beta(G_M)=1$); \emph{multicyclic} ($v(M)\le r$, $\beta(G_M)\ge2$);
  \emph{large} ($v(M)\ge r+1$, connected; coefficient at most $c_v\varrho_v^{-|M|}L$ by \Cref{lem:e:uniform}). Then
  $\Var(\Pi_D)\le4\sum_{\text{classes}}\Var(\Pi_{\text{class}})$.

  For two placed motifs $M,M'$ of one class let $X_M:=\eps^MB^{(r)}_M(Z)$. If their vertex sets are disjoint,
  $X_M$ and $X_{M'}$ are functions of $(Z_{V(M)},(A_e)_{e\in M})$ and $(Z_{V(M')},(A_e)_{e\in M'})$ and are
  independent. Otherwise $U:=M\uplus M'$ (multiplicities $a_e=m_e+m'_e$) is connected. Given $Z$, the pairs are
  independent, so by \Cref{lem:e:edge}(iii) and $C_\eps\Delta\le1$,
  \[
    |\Cov(X_M,X_{M'})|\le2(C_\eps\Delta)^{\sum_ec(a_e)}\|B_M\|_\infty\|B_{M'}\|_\infty\le2(C_\eps\Delta)^{2d(U)}\|B_M\|_\infty\|B_{M'}\|_\infty,
  \]
  where $|\Exp X_M\Exp X_{M'}|$ obeys the same bound because $c(m_e)+c(m'_e)\ge c(m_e+m'_e)$, and $c(a)\ge2$ for all
  $a\ge1$. Fix the isomorphism types of $M$, $M'$ and their overlap pattern (finitely many, depending only on
  $D$); there are at most $n^{v(U)}$ placements.
  \begin{itemize}[leftmargin=*]
    \item Tree class: $d(U)\ge d(M)=v(M)-1\ge m-1$ and $v(U)\le d(U)+1$, so
      $n^{v(U)}\Delta^{2d(U)}\le n\delta^{d(U)}\le n\delta^{m-1}=N_m$; the coefficients contribute at most $CL^2s^{2m}$.
    \item Unicyclic class: $U$ contains the cycle of $G_M$, so $\beta(U)\ge1$, $v(U)\le d(U)$ and $d(U)\ge3$; hence
      $n^{v(U)}\Delta^{2d(U)}\le\delta^{d(U)}\le\delta^3$; the coefficients contribute at most $CL^2s^2$.
    \item Multicyclic class: $\beta(U)\ge\beta(G_M)\ge2$, so $v(U)\le d(U)-1$, and a simple connected graph with
      cyclomatic number $2$ has at least $5$ edges; hence $n^{v(U)}\Delta^{2d(U)}\le\delta^{d(U)}/n\le\delta^5/n$; the
      coefficients contribute at most $CL^2$.
    \item Large class: $d(U)\ge v(M)-1\ge r$, so the placements contribute at most $n\delta^r$; the coefficients
      contribute at most $CL^2$.
  \end{itemize}
  Summing over the finitely many types gives the claim.
\end{proof}

\subsection{Remainder}

Let $\Rem_D:=\Sw_r-\Pi_D$, $\varsigma:=\sqrt\delta\log n$ and define the small-field event
$\mathcal G:=\{\max_{v\in[n],c\in[K]}|\sum_{u\ne v}\eps_{vu}\Ind{Z_u=c}|\le\varsigma\}$.

\begin{lemma}[Remainder]
  \label{lem:e:remainder}
  Assume \eqref{eq:e:A}, the conclusions of \Cref{lem:d:path}, $\delta\le1$ and $D\ge r+2$.
  \begin{enumerate}[label=(\roman*),leftmargin=*]
    \item There is $n_0$ depending only on $(q_0,w_{\mathrm{hi}})$ such that $P_t(\mathcal G^c)\le2Kn\exp(-q_0^2\log^2n/36)$
      for $n\ge n_0$.
    \item There are $C$, $n_R$ depending only on $(K,q_0,r,D,L,w_{\mathrm{hi}})$ and $c_R\in(0,1]$ depending only on
      $(K,r,w_{\mathrm{hi}})$ such that for $n\ge n_R$ and $\varsigma\le c_R/2$: on $\mathcal G$,
      $|\Rem_D|\le C\eta\,n(n\Delta)^{r+1}\varsigma^{D-r}$, and everywhere $|\Rem_D|\le Cn^{2D}$.
    \item Hence $\Exp_t\Rem_D^2\le C\big[n^{r+3}\delta^{D+1}(\log n)^{2(D-r)}+n^{4D+1}e^{-q_0^2\log^2n/36}\big]$.
  \end{enumerate}
\end{lemma}

\begin{proof}
  (i) Given $Z$, $\sum_u\eps_{vu}\Ind{Z_u=c}$ is a sum of at most $n$ independent terms of modulus at most
  $2\Delta/q_0$, with conditional mean of modulus at most $2w_{\mathrm{hi}}\delta/q_0$, which is at most $\varsigma/2$ for
  $n\ge n_0$, and conditional variance at most $4\delta/q_0^2$. Bernstein's inequality with deviation $\varsigma/2$,
  together with $\Delta\varsigma=\delta\log n/\sqrt n\le\delta$ for $n\ge n_0$, bounds the tail by
  $2\exp(-q_0^2\log^2n/36)$ for each $(v,c)$; take a union bound over the $nK$ pairs.

  (ii) \emph{Local decomposition.} By linearity,
  $\Sw_r=\sum_i\sum_{|S|\le r}\DS_S\Phi(f_i)+\sum_{i<j}\sum_{|S|\le r}\DS_S\chi_{ij}$. For a term with root set $R$
  ($R=\{i\}$ or $\{i,j\}$) put $\Omega:=R\cup S$, internal variables $x:=(\eps_e)_{e\subseteq\Omega}$ and external
  factors $\phi_{vc}:=\prod_{u\notin\Omega:z_u=c}(1+w_c\eps_{vu})$ for $v\in\Omega$. Given the labels outside $\Omega$, the
  term is an explicit function $g_{R,S}(x,\phi;z_\Omega)$: the conditional weight of $z'_U$, $U\subseteq S$, is
  proportional to $\prod_{k\in U}\theta_{z'_k}\phi_{k,z'_k}\prod_{e\subseteq\Omega,e\cap U\ne\emptyset}(1+g_e\eps_e)$,
  $f_{ic}=\phi_{ic}\prod_{k\in\Omega\setminus\{i\}:z_k=c}(1+w_c\eps_{ik})$, and $\chi_{ij}$ depends only on $z_i,z_j,\eps_{ij}$.

  \emph{Connected decomposition.} If the graph $(\Omega,\mathrm{supp}\,x)$ is disconnected, then $g_{R,S}=0$. Indeed,
  if $R=\{i,j\}$ is split between components then $x_{ij}=0$ and $\chi_{ij}=0$. Otherwise some component
  $\Omega_2$ avoids $R$, so $\Omega_2\subseteq S$; the external factors are fixed vectors that do not couple
  labels, so the conditional law factorizes between $\Omega_2$ and $\Omega\setminus\Omega_2$, the swept function does
  not depend on $z_{\Omega_2}$, and pairing $U$ with $U\triangle\{u\}$ for some $u\in\Omega_2$ as in
  \Cref{lem:e:local}(i) makes the alternating sum vanish. Inclusion--exclusion over edge sets therefore gives
  \[
    g_{R,S}=\sum_{B\subseteq E_\Omega:\ (\Omega,B)\ \text{connected}}x^Bh_B(x_B,\phi),\qquad
    h_B:=\int_{[0,1]^B}\partial_{x_B}g_{R,S}(\tau\odot x_B,0,\phi)\,\mathrm d\tau ,
  \]
  because the inclusion--exclusion difference of $g_{R,S}$ over $B$ vanishes when $(\Omega,B)$ is disconnected;
  in every surviving term, $|B|\ge s':=|\Omega|-1$.

  \emph{Radial continuation.} Replace all edge variables by $\lambda\eps$, $\lambda\in\Cpx$, and let
  $\mathcal R:=c_R/\varsigma\ge2$, with $c_R$ fixed below. On $\mathcal G$ and for $|\lambda|\le\mathcal R$:
  $|\lambda w_c\eps_{vu}|\le2c_Rw_{\mathrm{hi}}\Delta/(q_0\varsigma)=2c_Rw_{\mathrm{hi}}/(q_0\sqrt n\log n)\le1/2$ for $n\ge n_R$,
  so
  \[
    \log\phi_{vc}(\lambda)=\lambda w_c\sum_{u\notin\Omega}\eps_{vu}\Ind{z_u=c}+\rho_{vc},\qquad
    |\rho_{vc}|\le\sum_u|\lambda w_c\eps_{vu}|^2\le\frac{4c_R^2w_{\mathrm{hi}}^2}{q_0^2\log^2n}.
  \]
  The at most $r+2$ terms inside $\Omega$ change the full sum of $\mathcal G$ by at most $2(r+2)\Delta/q_0\le\varsigma$
  for $n\ge n_R$, so the linear part has modulus at most $2c_Rw_{\mathrm{hi}}$, hence $|\phi_{vc}(\lambda)-1|\le
  e^{3c_Rw_{\mathrm{hi}}}-1$, while the internal variables satisfy $|\lambda x_e|\le2c_R\Delta/(q_0\varsigma)\to0$.
  Choose $c_R$, depending only on $(K,r,w_{\mathrm{hi}})$, so small that every normalizer has modulus at least $1/2$.
  Then $g_{R,S}$ is analytic with modulus at most $C\eta$ on a fixed complex neighborhood of the relevant
  $(x,\phi)$, by the argument of \Cref{lem:e:uniform} ($|\Phi|\le\|\dot\theta\|_1$, $|\chi|\le\|\dot w\|_\infty/w_{\mathrm{hi}}$), and
  by Cauchy estimates, $|h_B(\lambda x_B,\phi(\lambda))|\le C'\eta$ for $|\lambda|\le\mathcal R$.

  \emph{Taylor tail.} $x^B$ is a monomial of degree $|B|$, so the Taylor polynomial of degree $D$ of
  $x^Bh_B$ in $\eps$ is $x^BT_{D-|B|}h_B$, with $T_j$ the Taylor polynomial of degree $j$. For $G$ analytic
  on $|\lambda|\le\mathcal R$, the homogeneous part of degree $j$ of $G(\eps)$ is $[\lambda^j]G(\lambda\eps)$, which has
  modulus at most $\sup_{|\lambda|=\mathcal R}|G|\cdot\mathcal R^{-j}$; since $\mathcal R\ge2$,
  we have $|(h_B-T_{D-|B|}h_B)(\eps)|\le2C'\eta\mathcal R^{-(D+1-|B|)}$. Moreover $|x^B|\le(2\Delta/q_0)^{|B|}$ and
  $2\Delta/q_0\le\varsigma/c_R$ for $n\ge n_R$. Whether $|B|\le D$ or $|B|\ge D+1$ (then $T=0$),
  \[
    |x^B(h_B-T_{D-|B|}h_B)|\le C\eta\,\Delta^{s'}\varsigma^{D+1-s'}.
  \]

  \emph{Counting.} For each $s'\le r$ there are at most $2n^{s'+1}$ pairs $(i,S)$ with $|\{i\}\cup S|=s'+1$, and for
  $1\le s'\le r+1$ at most $4n^{s'+1}$ triples $(i<j,S)$ with $|\{i,j\}\cup S|=s'+1$; each term has at most
  $2^{\binom{r+3}2}$ sets $B$. Since $n\Delta/\varsigma=\sqrt n/\log n\ge1$, the largest $s'$ dominates:
  \[
    |\Rem_D|\le C\eta\sum_{s'=0}^{r+1}n^{s'+1}\Delta^{s'}\varsigma^{D+1-s'}\le C''\eta\,n(n\Delta)^{r+1}\varsigma^{D-r}.
  \]

  \emph{Everywhere.} For real $\eps$, $|\DS_S\Phi(f_i)|\le2^r\max_c|\ell_c|\le2^rL$ and $|\DS_S\chi_{ij}|\le2^r\cdot4L\Delta/q_0$,
  so $|\Sw_r|\le Cn^{r+2}$; there are at most $n^{2D}$ motifs with $|M|\le D$, $|\eps^M|\le1$, and their coefficients
  are bounded by \Cref{lem:e:uniform}, so $|\Pi_D|\le Cn^{2D}$; as $D\ge r+2$, we have $|\Rem_D|\le Cn^{2D}$.

  (iii) $n^2(n\Delta)^{2r+2}\varsigma^{2(D-r)}=n^{r+3}\delta^{r+1}\cdot\delta^{D-r}(\log n)^{2(D-r)}$; on $\mathcal G^c$ use the
  bound everywhere and (i); finally $\eta\le L$.
\end{proof}

\subsection{The Information Inequality}

\begin{lemma}[Information Inequality]
  \label{lem:e:info}
  Let $r\ge1$, $D\ge r+2$, and take the path of \Cref{lem:d:path} with $0<s\le\min\{s_P,1\}$. There are $C_I$,
  $n_I$ and $\varsigma_0>0$ depending only on $(K,w,x,q_0,r,D)$ such that for $n\ge n_I$, $q\in[q_0,1-q_0]$,
  $\delta\le1$ and $\varsigma=\sqrt\delta\log n\le\varsigma_0$, for all $t\in[0,1]$,
  \[
    I(t)\le C_I\Big[N_ms^{2m}+\delta^3s^2+\frac{\delta^5}n+n\delta^r+n^{r+3}\delta^{D+1}(\log n)^{2(D-r)}+n^{4D+1}e^{-q_0^2\log^2n/36}\Big].
  \]
\end{lemma}

\begin{proof}
  For $n\ge n_I$, $\Delta=\sqrt{\delta/n}\le n^{-1/2}\le\Delta_0$, so \eqref{eq:e:A} holds; $\varsigma_0:=c_R/2$ meets
  the condition of \Cref{lem:e:remainder}. By \Cref{lem:a:sweep,lem:e:glauber}, $I(t)\le\Exp_t\Sw_r^2$ and
  $\Exp_t\Sw_r=\Exp_t\Sobs=0$, so $\Exp\Pi_D=-\Exp\Rem_D$ and
  $\Exp\Sw_r^2\le2\Exp\Pi_D^2+2\Exp\Rem_D^2=2\Var(\Pi_D)+2(\Exp\Rem_D)^2+2\Exp\Rem_D^2\le2\Var(\Pi_D)+4\Exp\Rem_D^2$.
  Insert \Cref{lem:e:pairing,lem:e:remainder}.
\end{proof}

\section{Proof of the Lower Bound and of the Consistency Threshold}
\label{app:lower}

\begin{proof}[Proof of \Cref{thm:main}(ii)]
  Fix $\kappa\in(0,1)$ and let $r:=\lfloor1/\kappa\rfloor+1$, so that $\kappa r>1$, and
  $D:=\max\{r+2,\lceil(r+4)/\kappa\rceil\}$. Take the path of \Cref{lem:d:path} for a witness $x$ of
  \Cref{cond:W}, let $s_1:=\min\{s_P,s_H,1\}$, let $C_I$ be the constant of \Cref{lem:e:info} for this pair
  $(r,D)$, $a_0:=\min\{1,(8C_I)^{-1/(2m)}\}$, and choose the amplitude
  \[
    s:=\min\{s_1,\ a_0N_m^{-1/(2m)}\}.
  \]
  The window $\delta\le n^{-\kappa}$ gives $\delta\le1$, and for $n$ large, $\varsigma\le n^{-\kappa/2}\log n\le\varsigma_0$.
  Term by term in \Cref{lem:e:info}:
  \begin{itemize}[leftmargin=*]
    \item $C_IN_ms^{2m}\le C_Ia_0^{2m}\le1/8$;
    \item $C_I\delta^3s^2\le C_In^{-3\kappa}\to0$ and $C_I\delta^5/n\to0$;
    \item $C_In\delta^r\le C_In^{1-\kappa r}\to0$;
    \item $C_In^{r+3}\delta^{D+1}(\log n)^{2(D-r)}\le C_In^{r+3-\kappa(D+1)}(\log n)^{2D}\to0$, because $\kappa(D+1)\ge r+4+\kappa$;
    \item $C_In^{4D+1}e^{-q_0^2\log^2n/36}\to0$.
  \end{itemize}
  Hence there is $n_L$ such that $\sup_{t\in[0,1]}I(t)\le1/4$ for $n\ge n_L$. Every $P_t$ on the path is a
  positive distribution on a finite set ($q+\Delta w_k(t)\in(0,1)$ because $\Delta w_{\mathrm{hi}}\le q_0/8$, and
  $\theta_k(t)>0$), and $P_t(A)$ is a polynomial in $(\theta(t),w(t))$, hence continuously differentiable in $t$.
  \Cref{lem:a:hellinger} gives $\TV(P_0,P_1)\le\frac12(1/4)^{1/2}=1/4$. By \Cref{lem:d:path}(v), $P_0$ and $P_1$ are
  the laws of the verdicts under $\theta^+$ and $\theta^-$ with the known sensitivities $w$, and by
  \Cref{lem:d:path}(vi) their entropies differ by at least $s|\ell(x)|$. \Cref{lem:a:lecam} yields
  \[
    R^*_n(w)\ge\frac{s^2\ell(x)^2}{16}=\frac{\ell(x)^2}{16}\min\{s_1^2,\ a_0^2N_m^{-1/m}\}\ge c_L\min\{1,N_m^{-1/m}\}
  \]
  with $c_L:=\ell(x)^2\min\{s_1^2,a_0^2\}/16$.
  The amplitude equals $s_1$ exactly when $N_m\le(a_0/s_1)^{2m}$, which is the range of constant risk; the
  estimates hold for every value of $N_m$. Since $r$ and $D$ depend only on $\kappa$, $s_1$ only on $(K,w,x)$,
  $L$ only on $(K,w,x,r)$, and $C_I$ only on $(K,w,x,q_0,r,D)$, the constants $c_L$ and $n_L$ depend only on
  $(K,w,x,q_0,\kappa)$; fixing one witness $x=x(w)$ for each $w$ makes them depend only on $(K,w,q_0,\kappa)$.
  \Cref{cond:W} is the only property of $w$ used, so this part does not need $\gap(w)>0$.
\end{proof}

\begin{proof}[Proof of \Cref{cor:threshold}]
  \emph{Sufficiency.} If $N_m\to\infty$, \Cref{thm:main}(i), which holds for every admissible $(q_n,\Delta_n)$,
  gives $\sup_\theta\Exp_\theta(\hH_n-H(\theta))^2\le C_U\max\{1/n,1/N_m\}^{1/m}\to0$.

  \emph{Necessity.} If $N_m=n\delta^{m-1}$ does not tend to infinity, it stays bounded by some $C$ along a
  subsequence, on which $\delta\le(C/n)^{1/(m-1)}\le n^{-1/(2(m-1))}=n^{-1/(4K-4)}$ for $n$ large. Along it
  \Cref{thm:main}(ii) with $\kappa=1/(4K-4)$ gives $R^*_n(w)\ge c_L\min\{1,C^{-1/m}\}>0$, so no sequence of
  estimators is uniformly consistent.

  For $\Delta_n=n^{-\beta}$, $N_m=n^{m-(2m-2)\beta}\to\infty$ if and only if $\beta<m/(2m-2)=(2K-1)/(4K-4)$. For a
  homogeneous judge on $K'$ meanings the threshold is $K'/(2K'-2)$ \citep[Cor.~1]{anonymous2026rarity}, which equals $(2K-1)/(4K-4)$ for $K'=2K-1$, and
  $K/(2K-2)-(2K-1)/(4K-4)=1/(4K-4)$.
\end{proof}

\section{Genericity of the Witness Condition}
\label{app:generic}

This appendix proves \Cref{prop:generic}. Let $K\ge3$, $a_c=1/w_c$, and consider the odd moment system
\begin{equation}
  \label{eq:g:E}
  X_{2i-1}(x)=\sum_ca_cx_c^{2i-1}=0,\qquad i=1,\dots,K-1 .
\end{equation}
For pairwise distinct $z_1,\dots,z_K$ let $\pi_c(z):=\prod_{d\ne c}(z_c-z_d)^{-1}$. These are the coefficients of the
divided difference of order $K-1$: $\sum_c\pi_c(z)z_c^j$ equals $0$ for $0\le j\le K-2$ and $1$ for $j=K-1$.

\begin{proposition}[Restatement of \Cref{prop:generic}]
  \label{prop:g:generic}
  Let $K\ge3$.
  \begin{enumerate}[label=(\alph*),leftmargin=*]
    \item If $\gap(w)>0$, every nonzero real solution $x$ of \eqref{eq:g:E} has pairwise distinct and positive
      squares $z_c:=x_c^2$, and $a_cx_c=\lambda\pi_c(z)$ for some $\lambda\ne0$. In particular its coordinates are
      pairwise distinct and nonzero, so under $\gap(w)>0$, \Cref{cond:W} holds if and only if some real solution
      has $\ell(x)\ne0$. Nonzero real solutions exist for every $w$ (\Cref{prop:sharp}).
    \item The set of $w\in(0,\infty)^K$ for which $\gap(w)=0$ or \Cref{cond:W} fails is Lebesgue-null.
    \item For $K=3$, if $\gap(w)>0$ and
      $g(w):=w_1^2\log^3(w_2/w_3)+w_2^2\log^3(w_3/w_1)+w_3^2\log^3(w_1/w_2)\ne0$, then \Cref{cond:W} holds. For
      example, $w=(1,2,4)$ has $\gap(w)=1/4$ and $g(w)=15\log^32$.
  \end{enumerate}
\end{proposition}

\subsection{Proof of (a)}
Write $\beta_c:=a_cx_c$ and $z_c:=x_c^2$, so that $X_{2i-1}(x)=\sum_c\beta_cz_c^{i-1}$. Let $\zeta_1<\dots<\zeta_p$ be the
distinct values of $z$, $G_k$ the classes with $z_c=\zeta_k$ and $B_k:=\sum_{c\in G_k}\beta_c$; then \eqref{eq:g:E} reads
$\sum_k\zeta_k^{i-1}B_k=0$ for $i=1,\dots,K-1$. If $p\le K-1$, the $(K-1)\times p$ Vandermonde matrix
$[\zeta_k^{i-1}]$ has full column rank, so every $B_k=0$. For a group with $\zeta_k>0$, $x_c=\epsilon_c\sqrt{\zeta_k}$ with
$\epsilon_c\in\{\pm1\}$, and $B_k=\sqrt{\zeta_k}\sum_{c\in G_k}\epsilon_ca_c=0$ says that the sums of $a_c$ over
$\{c\in G_k:\epsilon_c=1\}$ and over $\{c\in G_k:\epsilon_c=-1\}$ agree; these sets are disjoint and not both empty,
contradicting $\gap(w)>0$. So every group has $\zeta_k=0$ and $x=0$. A nonzero solution therefore has $p=K$:
the squares are distinct. The $(K-1)\times K$ Vandermonde matrix then has rank $K-1$ and its kernel is spanned
by $\pi(z)$, all of whose entries are nonzero, so $\beta=\lambda\pi(z)$ with $\lambda\ne0$ because $\beta\ne0$. If some
$z_c=0$ then $\beta_c=0\ne\lambda\pi_c(z)$, a contradiction; so all squares are positive. Distinct positive squares
give distinct nonzero coordinates. \qed

For $\gap(w)=0$ part (a) can fail: $a=(1,2,3)$ and $x=(1,1,-1)$ solve \eqref{eq:g:E} with a repeated coordinate.
\Cref{cond:W} itself may still hold when $\gap(w)=0$: for $a=(1,2,3)$, $x=(-1,4/5,-1/5)$ has distinct coordinates,
$X_1=X_3=0$ and $\ell(x)=\frac15\log(27/256)\ne0$.

\subsection{Parametrization}
Let $\Omega:=\{x\in\Real^K:x_c\ne0,\ x_c^2\text{ pairwise distinct},\ \alpha_c(x):=\pi_c(x^2)/x_c>0\text{ for all }c\}$.
The sign of $\pi_c(z)$ is $(-1)^{\#\{d:z_d>z_c\}}$, so $\alpha_c>0$ means $\mathrm{sign}\,x_c=(-1)^{\#\{d:x_d^2>x_c^2\}}$.
For each permutation $\sigma$ of $[K]$ let
\[
  \Omega_\sigma:=\{x:\ x_{\sigma(1)}^2<\dots<x_{\sigma(K)}^2,\ \mathrm{sign}\,x_{\sigma(j)}=(-1)^{K-j}\};
\]
$\Omega$ is the disjoint union of these $K!$ open cones. The signed square root maps the convex cone
$\{0<z_{\sigma(1)}<\dots<z_{\sigma(K)}\}$ diffeomorphically onto $\Omega_\sigma$, and the cone meets the hyperplane
$\sum_cz_c=1$ in a convex set, so $\Omega_\sigma$ and $\Omega_\sigma\cap S^{K-1}$ are connected. The coordinate permutation
$P_\sigma$ maps $\Omega_{\mathrm{id}}$ onto $\Omega_\sigma$. On $\Omega$ define the real-analytic function
\[
  \Lambda(x):=\sum_c\pi_c(x^2)\log\alpha_c(x).
\]
It is invariant under simultaneous permutation of coordinates, so $\Lambda\circ P_\sigma=\Lambda$ on $\Omega_{\mathrm{id}}$. For
$\tau>0$, $\pi_c(\tau^2z)=\tau^{-2(K-1)}\pi_c(z)$ and $\alpha_c(\tau x)=\tau^{-(2K-1)}\alpha_c(x)$, and $\sum_c\pi_c=0$ gives
$\Lambda(\tau x)=\tau^{-2(K-1)}\Lambda(x)$.

\emph{Relation to the system.} Let $\gap(w)>0$ and let $x$ be a nonzero real solution of \eqref{eq:g:E}. By (a),
$a_c=\lambda\alpha_c(x)$. If $\lambda>0$ then $x\in\Omega$; if $\lambda<0$, then $-x$ is also a solution and
$\alpha(-x)=-\alpha(x)$, so $a=|\lambda|\alpha(-x)$ with $-x\in\Omega$. In all cases we may take $x\in\Omega\cap S^{K-1}$ and $t>0$
with $a=t\alpha(x)$, and then, since $\alpha_cx_c=\pi_c(z)$ and $\sum_c\pi_c(z)=0$,
\begin{equation}
  \label{eq:g:L}
  \ell(x)=-t\sum_c\alpha_cx_c(\log t+\log\alpha_c)=-t\log t\sum_c\pi_c(z)-t\sum_c\pi_c(z)\log\alpha_c=-t\Lambda(x).
\end{equation}

\subsection{\texorpdfstring{$\Lambda$}{Lambda} Is Not Identically Zero}
For pairwise distinct $y\in\Real^K$ let $\Phi_K(y):=\sum_c\pi_c(y)\log|\pi_c(y)|$.

\begin{lemma}
  \label{lem:g:phi}
  For every $K\ge3$ there are pairwise distinct $y$ with $\Phi_K(y)\ne0$.
\end{lemma}

\begin{proof}
  For $K=3$ and $y=(0,1,2)$, $\pi(y)=(1/2,-1,1/2)$ and $\Phi_3(y)=-\log2$. Let $K\ge4$ and let $y'\in\Real^{K-1}$ have
  distinct coordinates with $\Phi_{K-1}(y')\ne0$; put $y=(y',T)$ with $T\to\infty$. For $c<K$,
  $\pi_c(y)=\pi_c(y')/(y'_c-T)$, and $\pi_K(y)=\prod_{d<K}(T-y'_d)^{-1}$. Split $\Phi_K(y)$ into three parts. First,
  $\sum_{c<K}\pi_c(y')\log|\pi_c(y')|/(y'_c-T)=-\Phi_{K-1}(y')/T+O(T^{-2})$. Second,
  $-\sum_{c<K}\pi_c(y')\log(T-y'_c)/(y'_c-T)=\sum_{c<K}\pi_c(y')\varphi(T-y'_c)$ with $\varphi(u):=(\log u)/u$; this is the
  divided difference of order $K-2$ of $y\mapsto\varphi(T-y)$ at $y'_1,\dots,y'_{K-1}$, which equals
  $(-1)^{K-2}\varphi^{(K-2)}(T-y_*)/(K-2)!$ for some $y_*$ in the convex hull of $y'$, and
  $\varphi^{(j)}(u)=O(u^{-j-1}\log u)$, so this part is $O(T^{-(K-1)}\log T)$. Third,
  $\pi_K\log|\pi_K|=O(T^{-(K-1)}\log T)$. For $K\ge4$ the last two parts are $o(T^{-1})$, so
  $T\Phi_K(y',T)\to-\Phi_{K-1}(y')\ne0$, and $\Phi_K(y',T)\ne0$ for large $T$.
\end{proof}

For even $K$, equally spaced nodes give $\Phi_K=0$, since $\pi_{K+1-c}=-\pi_c$; the lemma asserts only that
$\Phi_K$ is not identically zero.

\begin{lemma}
  \label{lem:g:lambda}
  For every $K\ge3$ and every $\sigma$, $\Lambda$ is not identically zero on $\Omega_\sigma\cap S^{K-1}$.
\end{lemma}

\begin{proof}
  By permutation invariance and homogeneity it suffices to find $x\in\Omega_{\mathrm{id}}$ with $\Lambda(x)\ne0$. Take $y$
  from \Cref{lem:g:phi}; $\Phi_K$ is permutation invariant, so let $y$ be increasing. For small $h>0$ let
  $z_c:=1+hy_c$ (increasing and positive) and $x_j:=(-1)^{K-j}\sqrt{z_j}$, so $x\in\Omega_{\mathrm{id}}$. On $\Omega$,
  $\alpha_c=|\pi_c(z)|/|x_c|$, hence
  \[
    \Lambda(x)=\sum_c\pi_c(z)\log|\pi_c(z)|-\tfrac12\sum_c\pi_c(z)\log z_c .
  \]
  The first sum equals $h^{-(K-1)}\Phi_K(y)$ exactly, because $\pi_c(z)=h^{-(K-1)}\pi_c(y)$ and $\sum_c\pi_c(y)=0$. The
  second is half the divided difference of order $K-1$ of $\log$ at $z_1,\dots,z_K$, which equals
  $\log^{(K-1)}(\zeta)/(K-1)!$ for some $\zeta\in[z_1,z_K]$ and stays bounded as $h\to0$. Hence
  $h^{K-1}\Lambda(x)\to\Phi_K(y)\ne0$.
\end{proof}

\subsection{Proof of (b)}
$\{\gap(w)=0\}$ is, in the coordinates $a$, a finite union of hyperplanes $\{\sum_{c\in I}a_c=\sum_{c\in J}a_c\}$, hence
null, and $a\mapsto w=1/a$ is a diffeomorphism of $(0,\infty)^K$, which preserves null sets. Let $\mathcal B$ be the
set of $a$ with $\gap>0$ for which \Cref{cond:W} fails. For $a\in\mathcal B$, \eqref{eq:g:E} has a unit real solution
(Borsuk--Ulam); as above we may take $x\in\Omega\cap S^{K-1}$ and $t>0$ with $a=t\alpha(x)$, and since \Cref{cond:W}
fails, (a) gives $\ell(x)=0$, i.e., $\Lambda(x)=0$ by \eqref{eq:g:L}. Therefore
\[
  \mathcal B\subseteq\Psi(\mathcal Z),\qquad\mathcal Z:=(0,\infty)\times\{x\in\Omega\cap S^{K-1}:\Lambda(x)=0\},\qquad\Psi(t,x):=t\alpha(x).
\]
On each connected component $\Omega_\sigma\cap S^{K-1}$ of the $(K-1)$-dimensional real-analytic manifold
$\Omega\cap S^{K-1}$, $\Lambda$ is real-analytic and not identically zero (\Cref{lem:g:lambda}), so its zero set is null \citep{mityagin2020zero}; hence $\mathcal Z$ is null in the $K$-dimensional manifold $(0,\infty)\times(\Omega\cap S^{K-1})$.
$\Psi$ is a smooth map between $K$-dimensional manifolds, Lipschitz on each of countably many coordinate
neighborhoods, so it maps null sets to null sets, and $\mathcal B$ is null. \qed

\subsection{Proof of (c)}
Let $K=3$ and $\gap(w)>0$. The $a_c$ are distinct, so $a$ and $a\odot\log a$ are linearly independent and the only
direction in the plane $X_1=0$ with $\ell=0$ is
$y:=a\times(a\odot\log a)=(a_2a_3\log(a_3/a_2),\ a_1a_3\log(a_1/a_3),\ a_1a_2\log(a_2/a_1))$. If \Cref{cond:W} fails,
every real solution of \eqref{eq:g:E}, and one exists by (a), has $\ell=0$ and is parallel to $y$, so $X_3(y)=0$. A
direct computation gives
\begin{align*}
  X_3(y)=\sum_ca_cy_c^3&=(a_1a_2a_3)^3\big[w_1^2\log^3(w_2/w_3)+w_2^2\log^3(w_3/w_1)+w_3^2\log^3(w_1/w_2)\big]\\
  &=(a_1a_2a_3)^3g(w),
\end{align*}
so $g(w)\ne0$ implies \Cref{cond:W}. The function $g$ is antisymmetric in the coordinates of $w$ and vanishes when
two of them coincide, in line with $\gap(w)=0$ there. For $w=(1,2,4)$, $a=(1,1/2,1/4)$, the smallest gap is
$|1-1/2-1/4|=|1/2-1/4|=1/4$, and $g(w)=(-\log2)^3+4(2\log2)^3+16(-\log2)^3=15\log^32$. \qed

\section{Numerical Checks}
\label{app:numerics}

The checks below test the mechanisms the proofs rely on and the size of their constants, and
\Cref{tab:numerics} summarizes them; the last subsection simulates the risk of the estimator. They run on a CPU with mpmath 1.3.0 \citep{mpmath2023mpmath} for
the high-precision computations, exact rational arithmetic where stated, and NumPy 1.26.4 \citep{harris2020array} elsewhere.

\begin{table}[!t]
  \centering
  \begin{threeparttable}
    \caption{Numerical checks of the proof mechanisms and constants.}
    \label{tab:numerics}
    \footnotesize
    \begin{tabular}{@{}>{\raggedright\arraybackslash}p{0.22\textwidth}>{\raggedright\arraybackslash}p{0.33\textwidth}>{\raggedright\arraybackslash}p{0.39\textwidth}@{}}
\toprule
check & setting & result \\
\midrule
Moment preservation and fifth-order scaling, $K=3$, $w_0$ & searched pair; $2001$-point grid, $80$ digits, six amplitudes $s\in[10^{-3},1]$ & relative drift $\max_t|S_v(t)-S_v(0)|/|S_v(0)|$ at most ${9.8\times10^{-81}}$ for $v\le4$; fitted order in $s$ of $\sup_t|\partial_tS_5|$: $5.0000$ \\ \addlinespace[3pt]
Exact Fisher information, ${K=3}$, three rational points (not $w_0$) & $9$ directions keeping $S_1,\dots,S_4$; $n\le7$; up to $\Delta^{12}$; exact arithmetic & in every direction: no term below $\Delta^6$; the $\Delta^6$ coefficient is $\binom n3\mu_2^3(\partial C_3)^2$; writing the $\Delta^k$ coefficient as $\sum_vc_{k,v}\binom nv$, $c_{k,v}=0$ for $v>k/2+1$; $c_{8,5}=125\mu_2^4(\partial S_5)^2$ \\ \addlinespace[3pt]
Endpoint $\chi^2$, $K=3$, $w_0$, ${q=1/3}$ & searched pair scaled toward $\theta^\star$ by ${s\in\{1,\frac12,\frac14,\frac18\}}$; $n\le7$; up to $\Delta^{12}$; $60$ digits & for $3\le n\le7$, every order below $\Delta^6$ at most ${3.2\times10^{-60}}$ relative to $\Delta^6$; for $6\le k\le12$, $|c_{k,v}|$ with $v>k/2+1$ at most ${7.0\times10^{-57}}$ relative to $\max_v|c_{k,v}|$; $c_{8,5}=125\mu_2^4(S_5^+-S_5^-)^2$ up to a relative error of at most ${4.4\times10^{-50}}$ \\ \addlinespace[3pt]
Modulus, $K=3$, $w_0$ & $200{,}000$ random pairs and an adversarial search & $|\Delta H|/\epsilon^{1/5}$ at most $2.29$ over the pairs found, versus $C_K(w_0)\approx3.0\times10^{4}$; lemma checks: no violation \\ \addlinespace[3pt]
Variance of $\hS_v$, $K=3$, $w_0$, $q=1/3$ & $n\in\{16,32\}$, $n\Delta^2\in\{0.05,0.2,1,2\}$, two distributions, $v=2,\dots,5$: $64$ tests; $20{,}000$ graphs each & ratio of the empirical variance to the bound of \Cref{lem:var} at most $0.0148$; ratio to the leading term $d_v\rho_q^{v-1}/(n)_v$ between $1.01$ and $1.62$ for $n\Delta^2\le0.2$ \\ \addlinespace[3pt]
Condition W at $w_0$ & exact rational arithmetic; exponential series with explicit tail bounds & $\ell(x)<-0.19$ for the witness; $\gamma(w_0)>0.29$; distinct coordinates \\ \addlinespace[3pt]
Reflected path, $K=3$, $w_0$ & $2001$-point grid, $50$ digits, sampled extrema & atom gaps at least $0.853$; $w(t)$ within $[0.453,1.448]$; entropy gap over $2s\ell(x)$: $1.0000000$ at $s=0.003$ \\ \addlinespace[3pt]
Genericity, $K=3,4,5$ & $2000$, $400$, $150$ random $w$; real zeros with $\|x\|_2=1$ found by multistart & squares distinct, $a\propto\alpha(x)$; in every draw the maximum of $|\ell(x)|/\|a\|_1$ over the zeros found is at least ${9.1\times10^{-7}}$, ${1.2\times10^{-3}}$, ${3.0\times10^{-4}}$ \\ \addlinespace[3pt]
\Cref{fig:path} path, $K=3$, ${w=(1,2,4)}$ & $201$-point grid, $40$ digits, sampled extrema & $w_k(t)$ within $[1.000,4.000]$; atom gaps at least $0.850$ \\
\bottomrule
\end{tabular}

    \begin{tablenotes}[flushleft]\footnotesize
    \item[]
      Bounds over the sampled points or the runs are rounded outward to the displayed precision, and other values
      to the nearest displayed digit. The triangle cumulant is $C_3:=\sum_c\theta_c^3w_c^3-3S_2S_3+2S_2^3$, the coefficient $C_M$ of
      \Cref{lem:e:exact} for the triangle. $\mu_2:=1/(q(1-q))$, $\partial C_3$ and $\partial S_5$ are the derivatives
      of the triangle cumulant and of the fifth tree moment, and $S_5^\pm$ are the fifth tree moments at the two
      endpoints.
    \end{tablenotes}
  \end{threeparttable}
\end{table}

\subsection{Sensitivity Profiles}
The $K=3$ checks use
\[
  w_0=(0.8187472594944942,\ 0.4532105734559928,\ 1.4478413485015227),
\]
whose coordinates are double-precision numbers, with $\gap(w_0)\approx0.294$. The exact Fisher expansion uses three rational parameter points $(q;\theta;w)$:
\[
  \big(\tfrac13;\,(\tfrac12,\tfrac3{10},\tfrac15);\,(\tfrac45,\tfrac12,\tfrac75)\big),\qquad
  \big(\tfrac25;\,(\tfrac25,\tfrac7{20},\tfrac14);\,(\tfrac32,\tfrac23,1)\big),\qquad
  \big(\tfrac14;\,(\tfrac13,\tfrac12,\tfrac16);\,(\tfrac12,\tfrac65,\tfrac9{10})\big).
\]

\subsection{Moment Preservation and Fifth-Order Scaling}
A pair at $w_0$ sharing $S_1,\dots,S_4$ is found by numerical search (their $80$-digit values are in the
repository):
\[
  \theta\approx(0.616019,\ 0.369993,\ 0.013988),\qquad\theta'\approx(0.423587,\ 0.196884,\ 0.379529).
\]
Both
endpoints are scaled toward $\theta^\star$ by an amplitude $s$: $\theta_c\mapsto a_c(\mu+s(w_c\theta_c-\mu))$, which keeps
$S_1,\dots,S_4$ equal. The scaled pair is joined by the path of \Cref{lem:d:path} built on its own moment sequence. Along this path the lower tree moments do not drift, and
$\sup_t|\partial_tS_5|$ scales as $s^5$, the order used in \Cref{lem:e:small}.

\subsection{Exact Expansions}
For $n\le7$ we expand two quantities in powers of $\Delta$ up to $\Delta^{12}$: the Fisher information of the verdict graph along
nine directions that keep $S_1,\dots,S_4$ fixed to first order (exact rational arithmetic, all isomorphism
classes of graphs on at most $7$ vertices), and the $\chi^2$ divergence between the two endpoints of the searched
pair scaled toward $\theta^\star$ ($60$-digit arithmetic). At each of the three rational points the directions are
the two vectors of the reduced row echelon basis of the kernel of the Jacobian of $(S_1,\dots,S_4)$ with
respect to $(\theta,w)$, and the first vector plus twice the second.

Write the coefficient of $\Delta^k$ as $\sum_vc_{k,v}\binom nv$. Both expansions start
at $\Delta^6$ with the triangle term $c_{6,3}$, and up to $\Delta^8$ the only other nonzero coefficients are
$c_{7,3}$, $c_{8,3}$, $c_{8,4}$ and the five-vertex tree term $c_{8,5}$. Relative to $\binom n3\Delta^6$ the other
terms on three and four vertices are smaller by factors of order $\Delta$, $\Delta^2$ and $n\Delta^2$, while the tree
term is of order $n^2\Delta^2$, and along the endpoint family, $c_{8,5}$ scales as $s^{10}=s^{2m}$. This is the split
of \Cref{lem:e:pairing} into unicyclic motifs, of order $\delta^3s^2$, and trees, of order $N_ms^{2m}$.
No coefficient grows faster in $n$ than the number of vertices of a connected motif allows: $c_{k,v}=0$ for
$v>k/2+1$.

\subsection{Modulus and Variance}
The modulus checks draw $200{,}000$ pairs from five families (independent, locally perturbed, near the center,
near the boundary, with zero coordinates) and run Nelder--Mead searches from $400$ starts for each of three
ratios. No evaluated pair violates the structural
claims of \Cref{lem:b:sign,lem:b:moments,lem:b:assign,lem:b:boundary}, and the largest
ratio $|\Delta H|/\epsilon^{1/5}$ found is far below the explicit constant $C_K(w_0)$. The variance checks simulate
$20{,}000$ verdict graphs for each of $16$ configurations ($q=1/3$, $n\in\{16,32\}$, $n\Delta^2\in\{0.05,0.2,1,2\}$, the
center $\theta\propto1/w_0$ and $\theta=(0.5,0.3,0.2)$) and compare the empirical variance of $\hS_v$, $v=2,\dots,5$, with
the bound of \Cref{lem:var} and with the leading noise term $d_v\rho_q^{v-1}/(n)_v$, where $\rho_q:=q(1-q)/\Delta^2$,
$d_2=2$ and $d_v=(v-1)!$ for $v\ge3$. The ratio of the empirical variance to the bound is at most $0.0148$,
and for $n\Delta^2\le0.2$ the ratio to the leading term lies between $1.01$ and $1.62$.

\subsection{Witness, Path Constants and Genericity}
At $w_0$, \Cref{cond:W} is verified in exact rational arithmetic, with the exponential bounded by partial sums
of its series and explicit tails. The reflected path at $w_0$ is sampled on a grid in $50$-digit arithmetic,
so its extrema are sampled values. For the genericity check, real solutions of \eqref{eq:g:E} are
found by multistart root finding for random sensitivity profiles; the search need not find every solution, and
every solution it finds has distinct squares and $a\propto\alpha(x)$, as \Cref{prop:g:generic}(a) states. For
$K=3$, the smallest value over the draws of the maximum of $|\ell(x)|/\|a\|_1$ over the solutions found with
$\|x\|_2=1$ occurs at a draw with $\gap(w)\approx9\times10^{-5}$, two reciprocal sensitivities being almost equal,
where also $|g(w)|\approx10^{-4}$.

\subsection{Entropy Risk at Matched Budgets}
\begin{figure}[!t]
  \centering
  \includegraphics{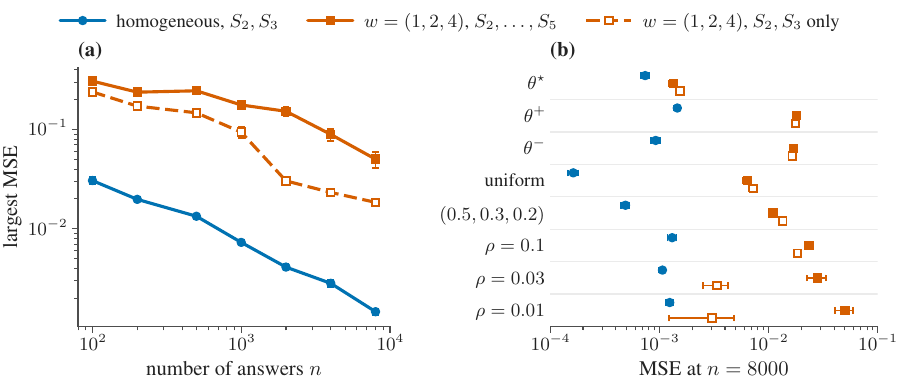}
  \caption{Mean squared error of the entropy estimate at matched budgets, $K=3$, $q=1/2$, $n\Delta^2=n^{-1/10}$: the
  homogeneous judge $w=\one$, projecting on $S_2,S_3$, and the judge $w=(1,2,4)$, projecting on $S_2,\dots,S_5$ or on
  $S_2,S_3$ only; $400$ simulated verdict graphs per point, bars one Monte Carlo standard error, in (a) that of the distribution with the largest error. (a) Largest error
  over the eight distributions of \Cref{tab:mse}. (b) Error at each distribution for $n=8000$: $\theta^\star\propto1/w$
  for $w=(1,2,4)$, its reflected pair $\theta^\pm$ at $s=0.4$, the uniform distribution, $(0.5,0.3,0.2)$, and
  $(1-2\rho,\rho,\rho)$.}
  \label{fig:mse}
\end{figure}

\Cref{fig:mse,tab:mse} report the mean squared error of the projected star-moment estimate of the entropy for $K=3$, $q=1/2$ and $n\Delta^2=n^{-1/10}$, a
fixed polynomial window, at $n\in\{100,200,500,1000,2000,4000,8000\}$, under two judges with the same $q$, $\Delta$ and
$n$: the homogeneous judge $w=\one$ and the judge $w=(1,2,4)$, which recognizes every meaning at least as well. The
projection uses $\hS_2,\dots,\hS_5$ under $w=(1,2,4)$, as in \Cref{alg:main}, and $\hS_2,\hS_3$ under $w=\one$, which determine the entropy
when $w=\one$; under $w=(1,2,4)$ the projection on $\hS_2,\hS_3$ alone is recorded as well.

The minimizer is found on
a grid of mesh $1/300$, refined locally to mesh $1/2400$. Each verdict graph is simulated exactly and enters only
through its degrees, since $X_{ij}$ takes one value on edges and another off them. Eight distributions are tested:
the center $\theta^\star\propto1/w$ of $w=(1,2,4)$, the reflected pair $\theta^\pm=a\odot(\mu\one\pm sx)$ of
\Cref{fig:simplex}(b) at $s=0.4$, the uniform distribution, $(0.5,0.3,0.2)$ and $(1-2\rho,\rho,\rho)$ for
$\rho\in\{0.1,0.03,0.01\}$, with $400$ graphs for each judge, distribution and $n$. The errors of the three projections are taken at these eight fixed distributions, so they are not minimax risks, and since the least-favorable family of the lower bound moves with $n$, the curves do not estimate the minimax rates.

\begin{table}[!t]
  \centering
  \begin{threeparttable}
    \caption{Mean squared error of the entropy estimate at matched budgets.}
    \label{tab:mse}
    \footnotesize
    \begin{tabular}{@{}lc*{5}{r}@{}}
\toprule
& & \multicolumn{2}{c}{$n=1000$} & \multicolumn{3}{c}{$n=8000$} \\
\cmidrule(lr){3-4}\cmidrule(l){5-7}
distribution $\theta$ & $\Ent(\theta)$ & $w=\one$ & $w=(1,2,4)$ & $w=\one$ & $w=(1,2,4)$ & $S_2,S_3$ only \\
\midrule
$\theta^\star\propto1/w$ & 0.956 & 3.92 & 3.31 & 0.74 & 1.34 & 1.55 \\
$\theta^+$, $s=0.4$ & 0.838 & 7.30 & 28.5 & 1.46 & 18.2 & 17.6 \\
$\theta^-$, $s=0.4$ & 0.896 & 5.13 & 26.6 & 0.93 & 17.1 & 16.5 \\
$(\frac13,\frac13,\frac13)$ & 1.099 & 1.02 & 11.2 & 0.16 & 6.40 & 7.25 \\
$(0.5,0.3,0.2)$ & 1.030 & 2.57 & 12.0 & 0.49 & 11.0 & 13.5 \\
$(0.8,0.1,0.1)$ & 0.639 & 6.10 & 40.4 & 1.31 & 23.6 & 18.5 \\
$(0.94,0.03,0.03)$ & 0.269 & 5.54 & 91.4 & 1.07 & 28.0 & 3.38 \\
$(0.98,0.01,0.01)$ & 0.112 & 6.01 & 177 & 1.25 & 50.3 & 3.05 \\
\midrule
largest &  & 7.30 & 177 & 1.46 & 50.3 & 18.5 \\
\bottomrule
\end{tabular}

    \begin{tablenotes}[flushleft]\footnotesize
    \item[]
      Mean squared errors are in units of $10^{-3}$, each from $400$ graphs; the last row gives the largest over the eight distributions. The projection uses $S_2,S_3$ under $w=\one$ and $S_2,\dots,S_5$ under $w=(1,2,4)$, and the last column gives the projection on $S_2,S_3$ only, under $w=(1,2,4)$. The Monte Carlo standard error is below $20\%$ of every mean squared error except those of the projection on $S_2,S_3$ only at $(0.94,0.03,0.03)$ and
      $(0.98,0.01,0.01)$, where it is $26\%$ and $60\%$.
    \end{tablenotes}
  \end{threeparttable}
\end{table}

Under $w=(1,2,4)$ the largest error over the eight distributions is between $10$ and $38$ times that of the homogeneous judge
at every $n$, for instance $0.177$ against $0.0073$ at $n=1000$ and $0.050$ against $0.0015$ at $n=8000$. The reflected
pair shows the mechanism of the lower bound: $\theta^+$ and $\theta^-$ share $S_2$, $S_3$ and $S_4$, while their fifth
moments differ by $1.8\times10^{-3}$ and their entropies by $0.058$. Under $w=(1,2,4)$ the error at $\theta^+$ falls from $0.038$ to $0.018$ as $n$ grows from $100$ to $8000$, against $0.027$ to $0.0015$ for the homogeneous
judge.

The largest errors occur near a vertex, where distributions far apart have close second tree moments:
$(0.17,0.55,0.28)$, with entropy $0.99$, has $S_2$ within $0.014$ of that of $(0.98,0.01,0.01)$, with entropy $0.11$,
while their $S_3$, $S_4$ and $S_5$ differ by $0.08$, $0.20$ and $0.34$. At $(0.98,0.01,0.01)$ the estimate is
off by more than $0.5$ for $45\%$ of the graphs at $n=100$ and $6\%$ at $n=8000$. For the homogeneous judge no simulated graph at that distribution gives such an error, and over all eight distributions such errors occur only at $n=100$, for at most $0.75\%$ of the graphs. Projecting on $\hS_2,\hS_3$ alone, which cannot determine
the entropy under $w=(1,2,4)$, lowers the error near the vertex but leaves the largest error between $7$ and $13$ times that of the homogeneous judge.

\end{document}